\documentclass[preprint,12pt]{elsarticle}

\usepackage{multirow}
\usepackage{pifont}

\usepackage{tikz}
\usepackage{caption}
\usepackage{subcaption}
\usepackage{amsmath, dsfont}
\usepackage{url}
\usepackage{booktabs}

\AtBeginDocument{%
  \providecommand\BibTeX{{%
    \normalfont B\kern-0.5em{\scshape i\kern-0.25em b}\kern-0.8em\TeX}}}

\begin{document}
\title{Does chronology matter in JIT defect prediction? 
A Partial Replication Study}

\author{Hadi Jahanshahi} 
\ead{hadi.jahanshahi@ryerson.ca}
\author{Dhanya Jothimani}
\author{Ay\c{s}e Ba\c{s}ar}
\author{Mucahit Cevik}
\address{Data Science Lab at Ryerson University, Toronto, ON M5B 1G3, Canada}

\begin{abstract}
 \textbf{BACKGROUND}: Just-In-Time (JIT) models, unlike the traditional defect prediction models, detect the fix-inducing changes (or defect inducing changes). These models are designed based on the assumption that past code change properties are similar to future ones. However, as the system evolves, the expertise of developers and/or the complexity of the system also change.   \\

\textbf{AIM}: In this work, we aim to investigate the effect of code change properties on JIT models over time. We also study the impact of using recent data as well as all available data on the performance of JIT models. Further, we analyze the effect of weighted sampling on the performance of fix-inducing properties of JIT models. For this purpose, we used datasets from four open-source projects, namely Eclipse JDT, Mozilla, Eclipse Platform, and PostgreSQL. \\

\textbf{METHOD}: We used five families of change code properties such as size, diffusion, history, experience, and purpose. We used Random Forest to train and test the JIT model and Brier Score (BS) and Area Under Curve (AUC) for performance measurement. We applied the Wilcoxon Signed Rank Test on the output to statistically validate whether the performance of JIT models improves using all the available data or the recent data.   \\

\textbf{RESULTS}: Our paper suggest that the predictive power of JIT models does not change by time. Furthermore, we observed that the chronology of data in JIT defect prediction models can be discarded by considering all the available data. On the other hand, the importance score of families of code change properties is found to oscillate over time.\\

\textbf{CONCLUSION}: To mitigate the impact of the evolution of code change properties, it is recommended to use weighted sampling approach in which more emphasis is placed upon the changes occurring closer to the current time. Moreover, since properties such as ``Expertise of the Developer" and ``Size" evolve with the time, the models obtained from old data may exhibit different characteristics compared to those employing the newer dataset. Hence, practitioners should constantly retrain JIT models to include fresh data.

\end{abstract}


\maketitle

\section{Introduction}\label{sec:introduction}
Software Quality Assurance (SQA) involves a set of standard activities such as inspection of codes and unit testing before the official release of a software system. These activities ensure that a software system meets the quality requirements and standards. However, limited SQA resources must be allocated wisely to minimize the risk and/or address the issues related to post-release defects. This led to the inception of \textit{software defect prediction models} \cite{li2006}.

Traditional approaches use historical data for training the models for defect prediction. This could be either at the method level \cite{giger2012, hata2012}, file level \cite{zimmermann2007} or subsystem level \cite{nagappan2005}. However, implementation of recommendation of the defect prediction models at such granularity becomes cumbersome in practice because of several reasons \cite{kamei2013, kamei2016}. Firstly, the likelihood of introduction of defects is higher in large files. Secondly, it may be difficult for a developer to recall the reasoning behind the implementation of a certain design decision over a time period. Lastly, the allocation of resources becomes difficult since many developers work on the same packages or the files \cite{kim2008}.

Researchers have proposed change-level defect prediction models (also known as Just-In-Time (JIT) models) to overcome the limitations of traditional defect prediction models. In JIT models, the likelihood of a code change introducing a defect/fix is predicted \cite{mockus2000, kamei2013, kim2008}. These models have the following advantages over traditional models: (a) Inspection of defects at change level is less cumbersome than at module level, (b) it is easier to allocate resources for handling the changes as the development of a module involves a group of authors or developers, and (c) since JIT models can track the changes, it enables the developers to inspect the problems as the design decisions would be relatively recent \cite{mcintosh2018, kamei2016}.

Despite several advantages of JIT prediction models, they require a large amount of historical data for improved model performance \cite{kamei2016}. Also, it works on the assumption that the properties of future events are similar to the properties of previous ones. However, this assumption may not hold true due to the dynamic nature of software development projects.  Hence, in this study, first, we investigated whether the properties of fix-inducing changes remain consistent with the evolution of the system over time, and second, we analyzed the importance of such a potential evolution in JIT defect prediction domain. For this purpose, as an initial step, we replicated the work by McIntosh and Kamei \cite{mcintosh2018}. Our objective behind this replication study is to validate the results from the previous study as well as understanding whether the results hold true for different datasets. A detailed discussion is provided in Section \ref{sec:motivationforthereplication}.  Our result using new dataset differs from the original study. Apart from addressing the first two research questions from their work, we formulated and addressed one more research question (RQ3). Accordingly, addressed research questions are as follows: 

(RQ1) \textit{Do JIT models lose predictive power over time?}

(RQ2) \textit{Does the relationship between code change properties and the likelihood of inducing a fix evolve?}

(RQ3) \textit{How can the performance of JIT models be improved while considering the chronology of the data?}




In this study, we used datasets from four open-source projects, namely, Eclipse JDP, Mozilla, Eclipse Platform and PostgreSQL \cite{kamei2013}. 

The contributions of our study are summarised as follows: 
\begin{itemize}
    \item Replication study helps in verifying the validity and reliability of the results of the original study. The process of replication leads to the development of new hypotheses and implementation of alternate techniques which are supposed to result in better interpretation. 
    \item We used a new dataset to check whether the obtained results can be generalised. 
    \item As an extension to the original study \cite{mcintosh2018}, we examined the performance of JIT models using all of the available data and the recent data (i.e.subset of all the data).  
    \item Additionally, we examined the effect of weighted sampling on the performance of JIT models. 
\end{itemize}

The organisation of the paper is as follows. Section \ref{sec:informationaboutoriginal} discusses the details related to the original study followed by information about the replication study in Section \ref{sec:informationaboutreplication}. Section \ref{sec:comparisonofresults} and Section \ref{sec:threatstovalidity} present the comparison of results with the original study and the threats to validity, respectively. Section \ref{sec:conclusion} concludes the paper.

\section{Information about the Original Study}\label{sec:informationaboutoriginal}
McIntosh and Kamei \cite{mcintosh2018} examined whether the properties of past events such as fix-inducing changes are similar to the properties of the future ones. To achieve this overall objective, the researchers formulated three research questions (see Table~\ref{tab:overviewofMcIntosh}). To address the research questions, six code properties, namely, size, diffusion, history, author and reviewer experiences, and review, were used for training JIT models on two open-source systems, i.e. {\scshape{Qt}} and {\scshape{OpenStack}}. These systems exhibited three characteristics such as traceability, rapidly evolving and code review policy. 

The study used the SZZ algorithm to check whether or not a change introduces a bug. To overcome the limitations of SZZ algorithm such as incomplete mapping of the fixing commit and bug and the systematic bias, the study considered the rates of fix-inducing changes and reviewed changes as well.

The code properties and the likelihood of bug were considered as the independent variables and dependent variables, respectively. During the data preprocessing phase, the study used Spearman rank correlation and variable clustering analysis to remove highly correlated code properties (i.e $\rho$ = 0.7). Further, the redundant change code properties were removed using the stepwise regression model. A nonlinear variant of the multiple regression model was fitted on the datasets. The accuracy of JIT models was analysed using the Area Under receiver operating characteristics Curve (AUC) and Brier Score (BS). 

To study how quickly JIT models lose their predictive power, the researchers carried out a longitudinal case study using two time horizons of three and six months for each project. Also, to understand how predictive power varies over time, they used short-period models and long-period models. In short-period models, JIT models were trained using the changes that occurred in one time period while, in long-period models, JIT models were trained using all the changes that occurred during or prior to a particular time period. Based on AUC and BS values, the performance of the models closer to the training period was found to be better than that of the models trained using the older periods. Also, \citet{mcintosh2018} deduced that long-period models do not retain predictive power longer than short-period models. 

To understand how the relationship between code change properties and the likelihood of inducing a fix evolves, the researchers used normalised Wald's $\chi^2$ score to compute the importance score of each family of code change properties. The analysis was carried out for both short-period and long-period models. In most cases, for both three-month and six-month time horizons, whether they are short-period or long-period models, ``Size" emerged as the most important family to predict the likelihood of bug. The importance of the code family ``Awareness" was low. In all cases, the importance of each family of code properties fluctuated over time indicating that the properties of fix-inducing changes tend to evolve. 

To examine how accurately the importance score of code properties in the current period represents the future ones, the researchers evaluated the stability of importance score for each family of code properties for short-period models and long-period models. The importance score of each family is computed during both training period (say, period $i$) and all the periods after the training period (say, period $j$). The difference in the importance score of each family in periods $i$ and $j$ was computed. If the difference is greater than zero, then the importance of the family is said to be overestimated during the training period $i$. Analysis indicated that the importance of the families such as ``Size" and ``Review" were consistently either overestimated or underestimated. This suggests that quality improvement plans should be revised periodically.

An overview of the research questions, dataset, and methodology is presented in Table \ref{tab:overviewofMcIntosh}. 

\begin{table*}[htbp]
\caption{Overview of the Original Study}
\begin{center}
\resizebox{\linewidth}{!}{
\begin{tabular}{|ll|}
\hline
\textbf{Motivation} & To analyse whether the properties of fix-inducing changes remain consistent as the systems evolve. \\ 
\textbf{Projects Used} & \scshape{Qt} and \scshape{OpenStack}\\
\textbf{Language} & R \\
\textbf{Data Extraction} & \multirow{2}{*}{}Database: VCS, ITS and code review\\ 
& Extraction of Code Properties: SZZ algorithm with additional analysis of the rates of fix-inducing changes and reviewed changes \\
\textbf{Time Period} & \multirow{2}{*}{}{\scshape{Qt}}: June 2011 to March 2014 \\
& {\scshape{OpenStack}}: November 2011 to February 2014\\ 
\textbf{Time Window} & 3 and 6 months \\
\textbf{Family of change code properties} & Size, Diffusion, History, Author/Reviewer Experience and Review\\ 
\textbf{Pre-processing}  & 
    \multirow{3}{*}{}Removal of highly correlated change properties using Spearman rank correlation ($\geq 0.7$) \\
    & Removal of redundant change properties \\ & Fitting the model using restricted cubic spline \\ 
\textbf{Model Used} & Logistic Regression \\ 
\textbf{Performance Measures} & AUC and BS \\
\textbf{Importance Measure} & Normalised Wald $\chi^2$ score \\
\textbf{Research Questions} & \multirow{3}{*}{}Do JIT models lose predictive power over time?  \\ 
 & Does the relationship between code change properties and the likelihood of inducing a fix evolve? \\ 
 & How accurately do current importance score of code change properties represent future ones?\\
\hline
\end{tabular}
}
\label{tab:overviewofMcIntosh}
\end{center}
\end{table*}

\section{Information about the Replication}\label{sec:informationaboutreplication}
\subsection{Motivation for Conducting the Replication}\label{sec:motivationforthereplication}
We conducted the replication study to mitigate the following limitations of the original study: 
\begin{enumerate}
    \item The original study was conducted on two open-source systems, namely, {\scshape{Qt}} and {\scshape{OpenStack}}. In order to generalise/validate results of the original study and to better understand the evolving nature of fix-inducing changes, we focus on historical datasets of other software systems, namely Eclipse JDT, Mozilla, Eclipse Platform, and PostgreSQL. 
    \item The results of the original study are specific to non-linear logistic regression. In our study, we used Random Forest (RF) to investigate whether the properties of fix-inducing changes remain consistent with the evolution of the system over time. 
    \item The original study used normalised Wald's $\chi^2$ score to compute the importance score of each family of code change properties. As mentioned in the original study, there is a need for defining a common importance score for other classifiers. Hence, we used accuracy-based measure and Gini index-based measure obtained using RF to determine the importance score of each family of the code change properties. 
\end{enumerate}

\subsection{Level of Interaction with the Original Researchers}\label{sec:levelofinteraction}
We carried out this study without any interaction with the original researchers. 

\subsection{Changes to the Original Study}\label{sec:changes}
We replicated the original study using the same dataset in original study (i.e., {\scshape{Qt}} and {\scshape{OpenStack}}). Since we obtained the same results as the original study, we did not report the results here\footnote{The results of the experiment using the same method and same dataset are available at https://github.com/HadiJahanshahi/JITChronology}.

As an extension to the original study, we introduced the following modifications:
\begin{enumerate}
    \item In our study, we defined one additional research question, which investigates different hypotheses to check the possibility of improvement in the performance of JIT models while considering the recency of the data.
    \item We used different datasets from four open-source projects, namely, Eclipse JDT, Mozilla, Eclipse Platform and PostgreSQL. 
    \item We considered a time window of only six months since the new datasets have a longer time horizon. We also reported the three-month time window analysis on GitHub.
    \item We used RF instead of Logistic Regression for prediction. We used RF as they are less biased and are less likely to overfit due to the random selection of a subset of features. Also, they are simple to interpret and implement, robust to outliers and can handle non-linearity in the data \cite{breiman2001, ren2015}. Moreover, previous studies have demonstrated that RF outperforms traditional predictive algorithms such as Logistic Regression and Decision Trees, for software defect prediction \cite{kamei2010, kamei2016}. Furthermore, the original study states that the findings of RQ1 are reproducible in the Random Forest context.
    \item In RQ2, we used accuracy-based measure and Gini index-based measure for computing the variable importance to study the evolution of the relationship between code change properties and the likelihood of inducing a fix since the Wald $\chi^2$ is not well investigated for Random Forest classifier. 
\end{enumerate}

\subsection{Data}\label{sec:data}
In our study, we used datasets from four open-source projects, namely, Eclipse JDT, Mozilla, Eclipse Platform, PostgreSQL. All the datasets have two characteristics mentioned in the original paper: traceability and code review policy. Five code properties, namely, Size \cite{mockus2000, kamei2013, kamei2016, mcintosh2018, kim2008}, Diffusion \cite{mcintosh2018, kamei2013, kamei2016, mockus2000}, History \cite{mcintosh2018, kamei2013, kamei2016}, Author/Reviewer Experience \cite{mcintosh2018, kamei2013, mockus2000, kamei2016} and Review \cite{mcintosh2018} are used in our analysis. Size represents the amount of code modified. Diffusion measures the variation of the change across a codebase. History represents the properties of a subsystem with changes. The expertise of an author or a reviewer is measured using the Author/Reviewer Experience. To examine the consistency in the properties of fix-inducing changes, we stratified the dataset into the time window of six months.

A summary of the dataset and an overview of families of code and review properties are presented in Table \ref{tab:summaryofdata} and Table  \ref{tab:summaryofcodeandreviewproperties}, respectively. 

\begin{table}[htbp]
\caption{Summary of the dataset}
\begin{center}
\resizebox{0.7\textwidth}{!}{\begin{tabular}{|l|l|l|}
\hline
{\textbf{Project Name}} & \textbf{Period} & \textbf{No. of Changes} \\
\hline
Eclipse JDT (JDT) & May 2001 - December 2007 & 35,386 (14\%) \\
Mozilla (MOZ) & Januay 2000 - December 2006 & 98,275 (5\%) \\
Eclipse Platform (PLA) & May 2001 - December 2007 & 64,250 (15\%) \\
PostgreSQL (POS) & July 1996 - May 2010 & 20,431 (25\%)
\\ \hline
\end{tabular}
}
\label{tab:summaryofdata}
\end{center}
\end{table}

\begin{table*}[htbp]
\scriptsize
\caption{Summary of Code and Review Properties \cite{kamei2016, kamei2013}}
\begin{center}
\resizebox{\linewidth}{!}{\begin{tabular}{|lp{2cm}p{3.1cm}p{5.4cm}|}
\hline
& \textbf{Property} & \textbf{Description} & \textbf{Rationale}  \\ \hline
\multirow{3}{*}{\rotatebox[origin=c]{90}{Size}} & Lines added (la) & The number of lines added to a code  & The more a code is modified, the higher are the chances for introduction of bugs \cite{nagappan2006, kamei2016, mcintosh2018} \\
& Lines deleted (ld) & The number of lines deleted from a code &    \\
& Lines of code before the change (lt) & Lines of code in a file before the change & Larger the file or module, higher is the likelihood of a defect \cite{koru2009} \\\hline
\multirow{4}{*}{\rotatebox[origin=c]{90}{Diffusion}} & Subsystems (ns) & The number of modified subsystems & Focused changes are less risky compared to scattered ones since the scattered changes require wide spectrum of expertise \cite{ambros2010, hasan2009}  \\
& Directories (nd) & The number of modified directories  &    \\
& Files (nf) & The number of modified files &   \\
& Entropy (Ent) &  The distribution of modified codes across files &   \\ \hline
\multirow{3}{*}{\rotatebox[origin=c]{90}{History}} & Unique changes (nuc) & The number of unique changes to the modified files &  The more the number of changes higher is the likelihood for defect as it requires the developers to recall and track previous modifications \cite{kamei2013}  \\
& Developers (ndev) & The number of developers who changed the modified files in the past & More the number of developers involved higher is the likelihood for defects \cite{matsumoto2010}   \\
& Age (age) & The amount of time between the last and current changes & Recently modified code is riskier compared to older code \cite{graves2000}   \\ \hline
\multirow{3}{*}{\rotatebox[origin=c]{90}{Experience}} & Developer Experience (exp) & Developer experience is measured using the number of prior changes made by a developer & Likelihood of risk is higher when a novice modifies a code when compared to the modifications made by an experienced developer \cite{mockus2000}  \\
& Recent changes (rexp) & The number of prior changes made by an actor weighted by age of modifications &   \\
& Subsystem changes (sexp) & The number of prior changes made by an actor in a subsystem  &   \\
 \hline
\multirow{1}{*}{\rotatebox[origin=c]{90}{Purpose}} & Fix & Whether a change fixes the defect or not & The likelihood of defect caused by the revisions that fix defects is higher than those changes that implement new functionality \cite{kamei2016, guo2010}\\
 
\hline
\end{tabular}}
\label{tab:summaryofcodeandreviewproperties}
\end{center}
\end{table*}


\subsection{Pre-processing, Model Building and Variable Importance}\label{sec:preprocessing}
In the pre-processing stage, we used Spearman rank correlation test\footnote{preferred over other tests as it does not require the data to be normally distributed.} to remove the code change properties that are highly correlated. We randomly selected one of the variables when two change properties have a correlation of 0.7 or more. Also, we checked for multicollinearity. 

We divided each dataset based on a time horizon of six months. Since the dataset was imbalanced, we performed undersampling in the training dataset.  

We trained and predicted the model using RF. RF is a machine learning algorithm used for both classification and regression problems. It works on the principle of building a number of decision trees during the training phase \cite{breiman2001}. A random subset of all the predictors is used for splitting each node in the decision tree. This ensures that the decision trees are not highly correlated. Since RF employs numerous decision trees, each may report different class. So, the final class is allocated by a majority vote of the classifiers.

To understand the importance of code change properties in predicting the likelihood of inducing a fix over time, we used the variable importance criteria obtained using RF. The two criteria for selecting the important variables are: (1) accuracy-based importance (referred as Type I measure in our study), which measures the percentage decrease in the accuracy when a variable is excluded and (2) Gini index-based importance (referred as Type II measure), which measures the decrease in impurity when a predictor is selected for splitting a node. Gini impurity index (G) is calculated as $G = \sum_{i=1}^{n_c}p_i(1-p_i)$, where $n_c$ is the number of classes in the target variable and $p_i$ is the ratio of this class. Both approaches have certain advantages. For example, the Gini index-based approach requires minimal computation as it is evaluated during the training stage, and also, it has a limitation of being biased when used for multi-class classification \cite{strobl2007}. Hence, we used both measures for evaluating the importance score of the predictors with the likelihood of inducing a fix. 

\subsection{Model Performance}\label{sec:modelperformance}
We used two threshold-independent performance measures, namely, the Area Under the receiver operator characteristics Curve (AUC) and Brier Score (BS), to determine the accuracy of JIT models. 

For AUC, we plot the true positive rate in the ordinate and false positive rate along the abscissa against various threshold values, which explains the discriminatory power of the model. In this case, AUC helps to understand whether the JIT model is able to correctly classify a change as fix-inducing. AUC values of 0, 0.5 and 1 represent the worst, random, and best discrimination, respectively.  

On the other hand, BS aids in measuring the calibration ability of the model. It is calculated using $BS = \frac{1}{N}\sum_{j=1}^{N}(x_j - \hat{x}_j)^2 $ where, $x_j = 1$ if the $j$th change is fix-inducing and $x_j = 0$ otherwise, $\hat{x}_j$ is the probability of the $j$th change as fix-inducing according to the model and $N$ represents the total number of changes. The lower the BS is, the better the calibration ability of the model will be \cite{mcintosh2018}. BS provides a better estimate of predicted probabilities in case of both balanced and imbalanced datasets \cite{hand2009}. It is preferred over AUC as it can be improved by systematically and consistently predicting probability values only using the best estimate \cite{gneiting2007, malley2012}.

\section{Comparison of Results with the Original Study}\label{sec:comparisonofresults}

\subsection{(RQ1) Do JIT Models Lose Predictive Power Over Time?}
\begin{figure}[!ht]
\centerline{\includegraphics[width=0.7\linewidth]{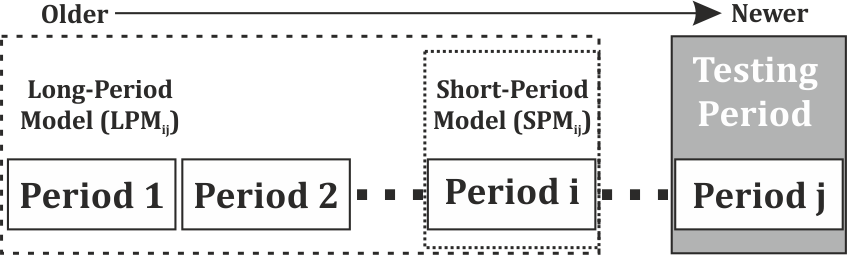}}
\caption{Long-Period vs. Short-Period models \cite{mcintosh2018}}
\label{fig:Short-Long}
\end{figure}

After dividing each dataset into six-month periods, we investigated how the performance of a JIT model evolves with time. For each time period, we trained two types of models:

\begin{enumerate}
    \item \textit{Short-Period models ($SPM$)}: $SPM_{ij}$ is trained on the changes that occurred in period $i$ and tested on the changes occurring in period $j$ while $i<j$. These models check whether the performance of an older dataset is comparable to that of a newer one (see Figure~\ref{fig:Short-Long}).
    \item \textit{Long-Period models ($LPM$)}: $LPM_{ij}$ is trained on all the changes that occurred from period $1$ to $i$ and tested on the changes occurring in period $j$ while $i<j$. Few studies suggest that larger amount of data results in better performance of JIT models \cite{Rahman:2013,kamei2016}. Therefore, unlike $SPM$ which considers the pairwise comparison of the data in two different time period, $LPM$ utilises the characteristics and information of all available training data until the time period $j$ (refer Figure~\ref{fig:Short-Long}).
\end{enumerate}

\begin{figure*}[!ht]
    \centering
        \begin{subfigure}[t]{0.5\textwidth}
        \centering
        \includegraphics[width=\textwidth]{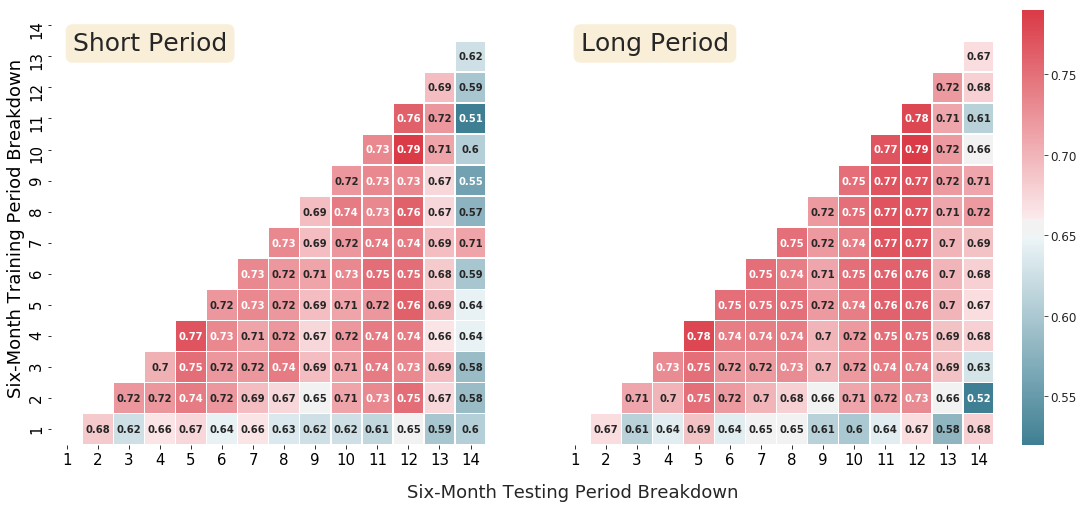}
        \vspace{-0.6cm}
        \caption{AUC in six-month periods ({\scshape JDT}).}
        \label{fig:RQ1-1a}
    \end{subfigure}%
    ~ 
    \begin{subfigure}[t]{0.5\textwidth}
        \centering
        \includegraphics[width=\textwidth]{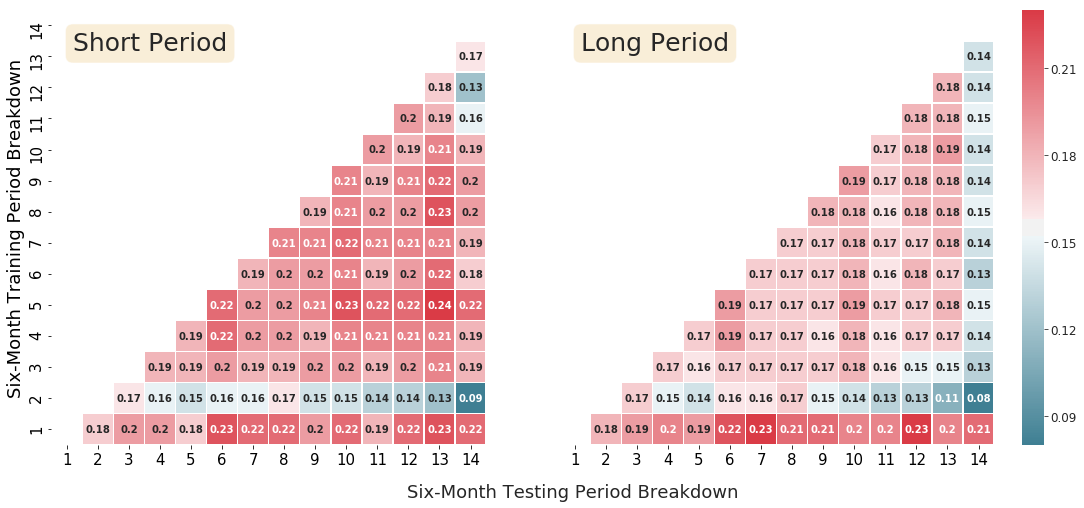}
        \vspace{-0.6cm}
        \caption{Brier in six-month periods ({\scshape JDT})}
        \label{fig:RQ1-1b}
    \end{subfigure}
    ~\\
        \begin{subfigure}[t]{0.5\textwidth}
        \centering
        \includegraphics[width=\textwidth]{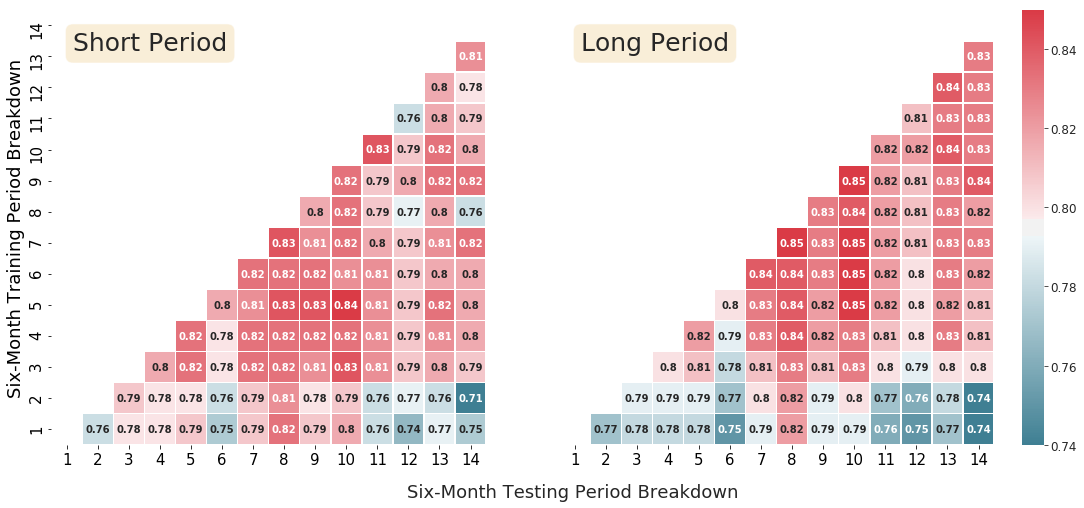}
        \vspace{-0.6cm}
        \caption{AUC in six-month periods ({\scshape Mozilla}).}
        \label{fig:RQ1-1c}
    \end{subfigure}%
    ~ 
    \begin{subfigure}[t]{0.5\textwidth}
        \centering
        \includegraphics[width=\textwidth]{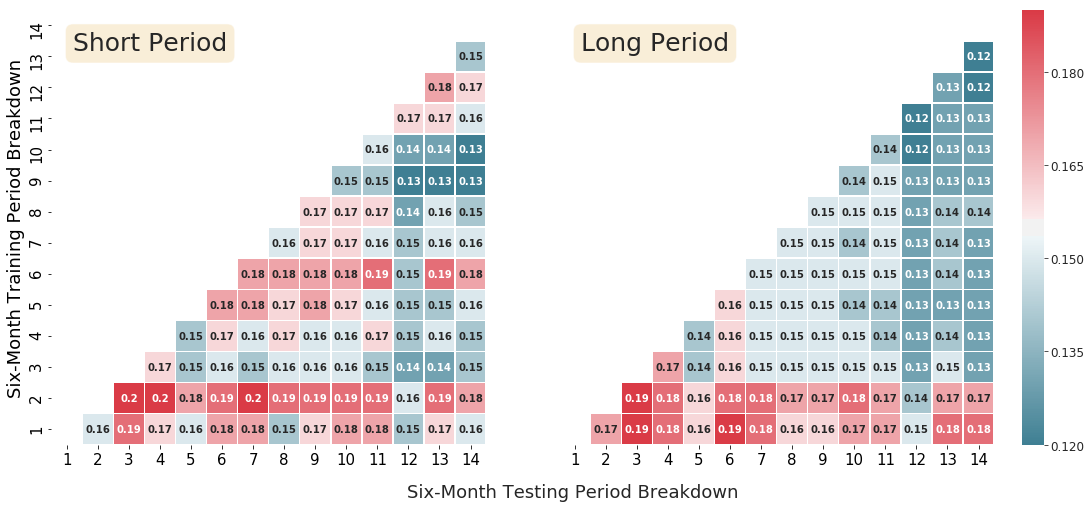}
        \caption{Brier in six-month periods ({\scshape Mozilla})}
        \label{fig:RQ1-1d}
    \end{subfigure}
    ~\\
        \begin{subfigure}[t]{0.5\textwidth}
        \centering
        \includegraphics[width=\textwidth]{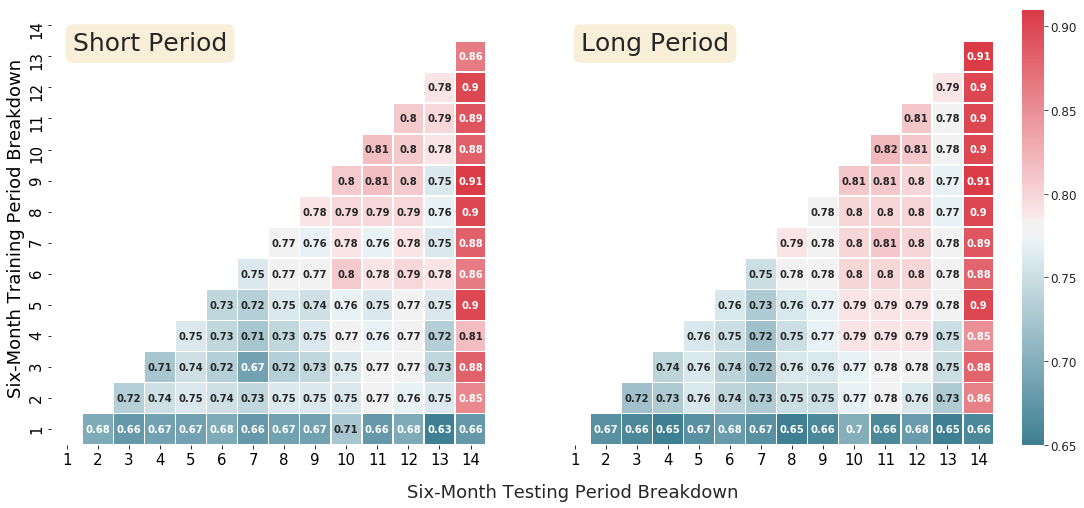}
        \vspace{-0.6cm}
        \caption{AUC in six-month periods ({\scshape Platform}).}
        \label{fig:RQ1-1e}
    \end{subfigure}%
    ~ 
    \begin{subfigure}[t]{0.5\textwidth}
        \centering
        \includegraphics[width=\textwidth]{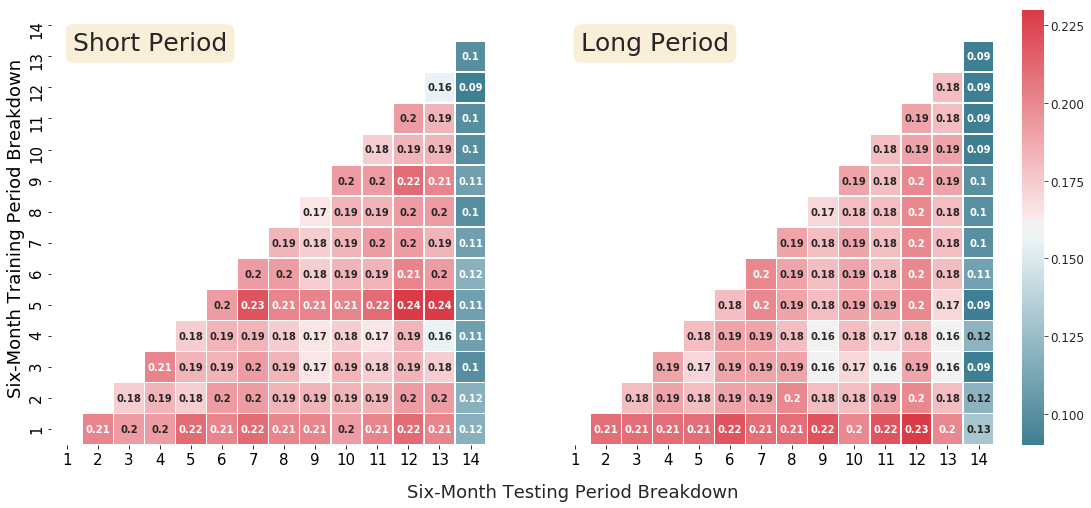}
        \vspace{-0.6cm}
        \caption{Brier in six-month periods ({\scshape Platform})}
        \label{fig:RQ1-1f}
    \end{subfigure}
    ~ \\
    \begin{subfigure}[t]{0.48\textwidth}
        \centering
        \includegraphics[width=\textwidth]{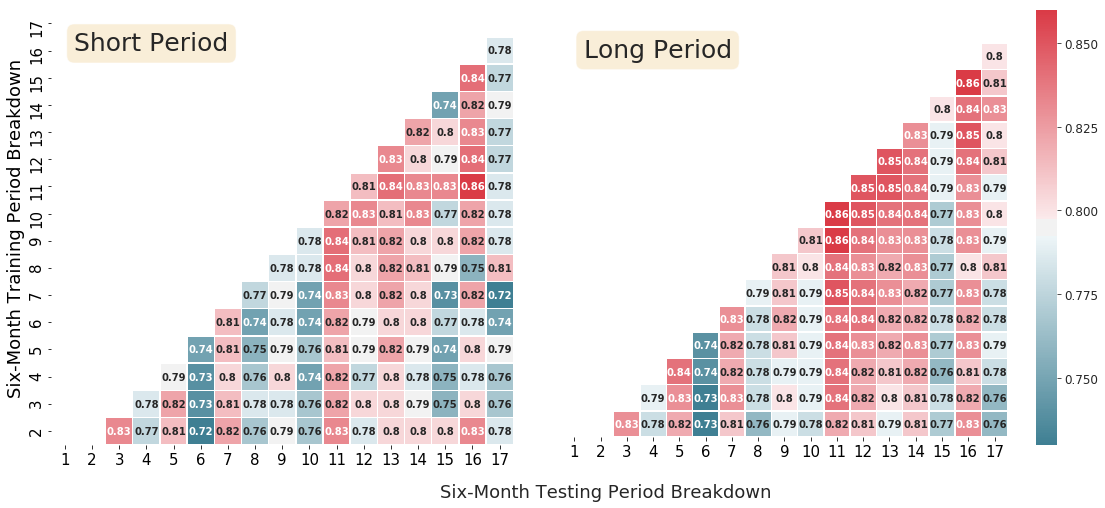}
        \vspace{-0.6cm}
        \caption{AUC in six-month periods ({\scshape Postgres}).}
        \label{fig:RQ1-1g}
    \end{subfigure}%
    ~ 
    \begin{subfigure}[t]{0.5\textwidth}
        \centering
        \includegraphics[width=\textwidth]{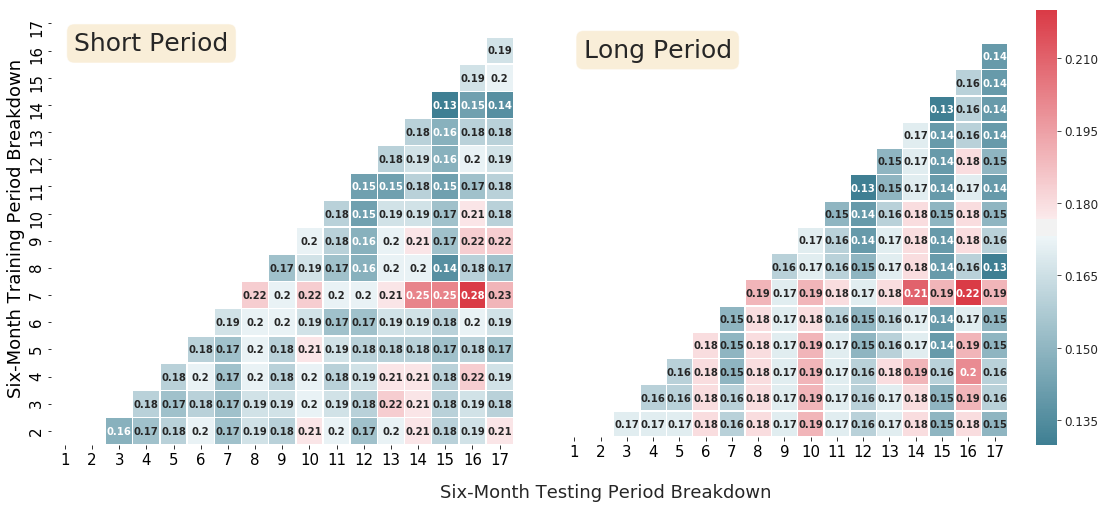}
        \vspace{-0.6cm}
        \caption{Brier in six-month periods ({\scshape Postgres})}
        \label{fig:RQ1-1h}
    \end{subfigure}

    \caption{The predictive performance of JIT systems based on the age of the training and testing set}
\label{fig:RQ1-1}
\end{figure*}

The performances of SPM-based and LPM-based JIT models are reported in terms of its discriminatory power (AUC) and calibration ability (BS). In Figure~\ref{fig:RQ1-1}, the entries along the diagonal and the upper diagonal are always empty since no model can be trained and tested in the same period or in a period which comes after the given time period. For example, in Figure~\ref{fig:RQ1-1a}, for short periods, $SPM_{27} = 0.69$ and $SPM_{67} = 0.73$. $SPM_{27}$ indicates AUC of the model when trained on time period $2$ and tested on time period $7$. On the other hand, in the same figure, for long periods, $LPM_{27} = 0.70$ and $LPM_{67} = 0.75$. $LPM_{27}$ indicates AUC of the model when trained on all data of time periods $i= 1, 2$ and tested on data from time period $ j = 7$. Furthermore, the performance of the model for six-month periods is illustrated using heatmaps, where higher AUC (red colour) and lower Brier (blue colour) are desirable.

In Figure~\ref{fig:RQ1-1}, 
unlike what we observed in the original study, there is no significant column-wise change in the figure. We expected to see the better performance of the model as we get closer to the diagonal of the matrix; however, for these datasets, the effect of changes in testing period is dominant over the changes in the training period.

Excluding the first training period breakdown, we were unable to observe any clear differentiation between old and new data. Although in most cases, the best values occur close to the diagonal of the matrix, the results obtained for RQ1 are not consistent with the original study in which a clear improvement in the performance has been observed. Since both the performance of the model and the evolution of the importance of each code change property are important, in RQ2, we will investigate the trade-off between the performance and feature importance of the models. 

The inconsistency of the result may be due to either the change in the classification method or the nature of the new dataset and metrics which are utilised in our paper. To investigate it further, we repeat the same Research Question using Logistic Regression to see if the inconsistency arose because of the classification method\footnote{The result and the scripts of this experiment are accessible at https://github.com/HadiJahanshahi/JITChronology}. The new design of the experiment produces the same result and reinforce the same conclusion that the interpretation of the original study can not be generalised to any of the new datasets. This issue has been discussed further in Section~\ref{sec:threatstovalidity}.

\noindent\fcolorbox{black}{white}{%
    \minipage[t]{\dimexpr1\linewidth-2\fboxsep-2\fboxrule\relax}
        \textit{In general, the predictive power of JIT models does not show any significant change as we move towards older data --- that is, the predictive performance of JIT models is not affected by the time.}
    \endminipage}

\subsection{(RQ2) Does the relationship between code change properties and the likelihood of inducing a fix evolve?}

\begin{figure*}[!ht]
    \centering
    \begin{subfigure}[t]{0.48\textwidth}
        \centering
        \includegraphics[width=\textwidth]{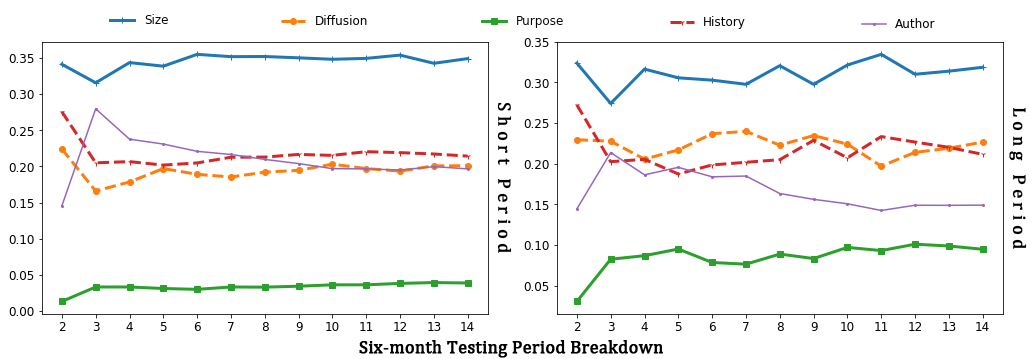}
        \caption{Six-month periods ({\scshape JDT}) - Type I}
        \label{fig:RQ2-1a}
    \end{subfigure}%
    ~ 
    \begin{subfigure}[t]{0.48\textwidth}
        \centering
        \includegraphics[width=\textwidth]{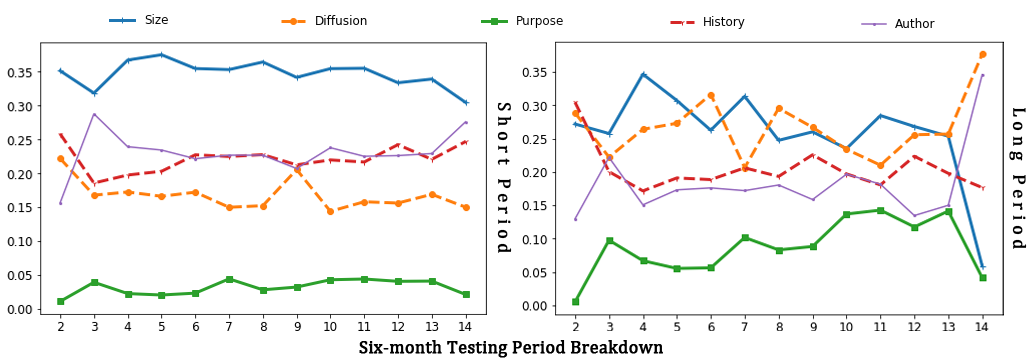}
        \caption{Six-month periods ({\scshape JDT}) - Type II}
        \label{fig:RQ2-1b}
    \end{subfigure}
    ~ \\
    \begin{subfigure}[t]{0.48\textwidth}
        \centering
        \includegraphics[width=\textwidth]{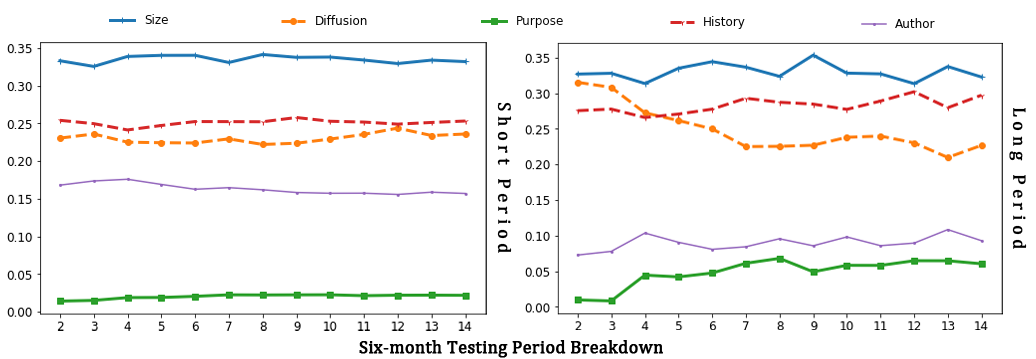}
        \caption{Six-month periods ({\scshape Mozilla}) - Type I}
        \label{fig:RQ2-1c}
    \end{subfigure}%
    ~ 
    \begin{subfigure}[t]{0.48\textwidth}
        \centering
        \includegraphics[width=\textwidth]{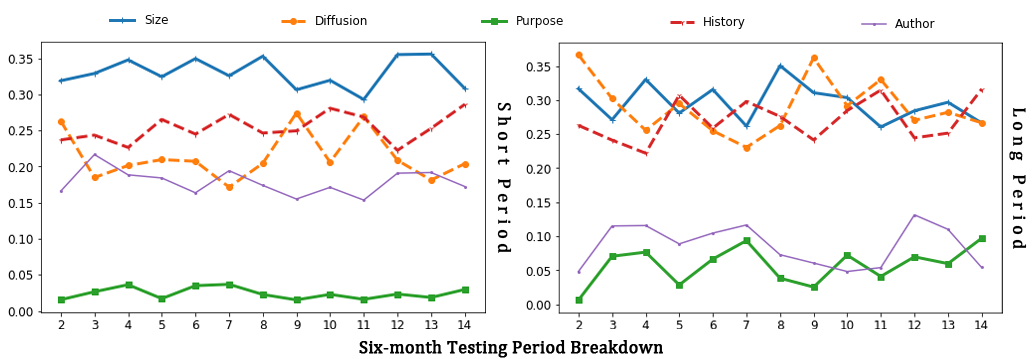}
        \caption{Six-month periods ({\scshape Mozilla}) - Type II}
        \label{fig:RQ2-1d}
    \end{subfigure}
        ~ \\
    \begin{subfigure}[t]{0.48\textwidth}
        \centering
        \includegraphics[width=\textwidth]{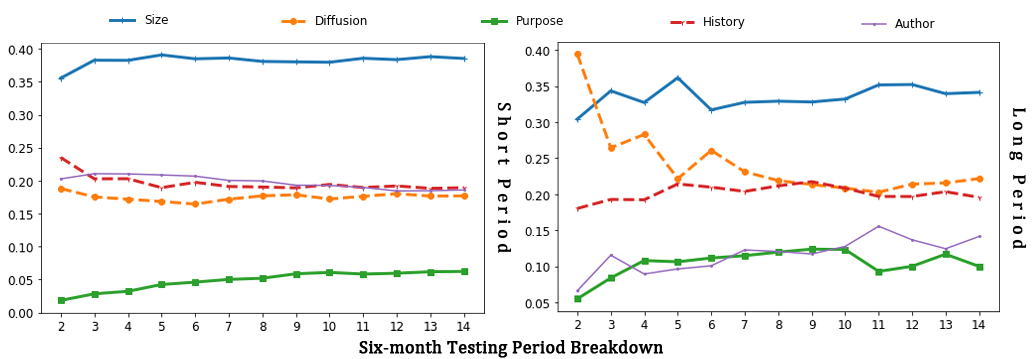}
        \caption{Six-month periods ({\scshape Platform}) - Type I}
        \label{fig:RQ2-1e}
    \end{subfigure}%
    ~ 
    \begin{subfigure}[t]{0.48\textwidth}
        \centering
        \includegraphics[width=\textwidth]{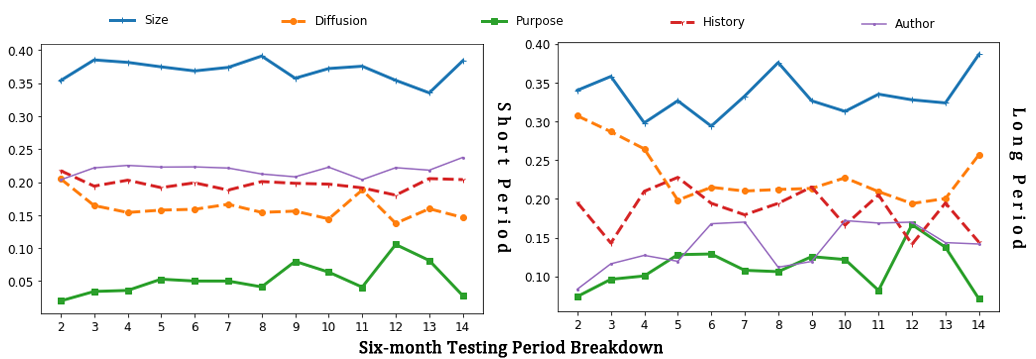}
        \caption{Six-month periods ({\scshape Platform}) - Type II}
        \label{fig:RQ2-1f}
    \end{subfigure}
        ~ \\
    \begin{subfigure}[t]{0.48\textwidth}
        \centering
        \includegraphics[width=\textwidth]{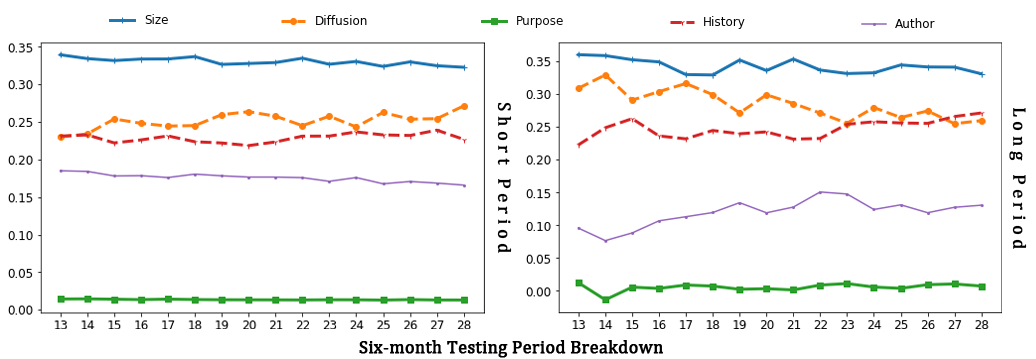}
        \caption{Six-month periods ({\scshape Postgres}) - Type I}
        \label{fig:RQ2-1g}
    \end{subfigure}%
    ~ 
    \begin{subfigure}[t]{0.48\textwidth}
        \centering
        \includegraphics[width=\textwidth]{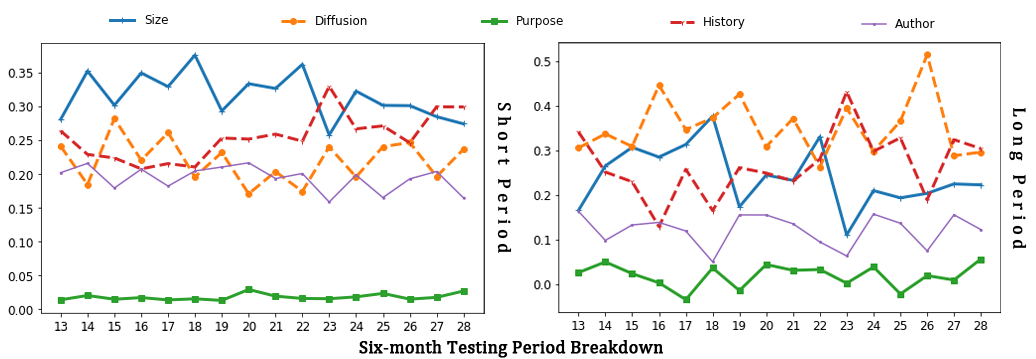}
        \caption{Six-month periods ({\scshape Postgres}) - Type II}
        \label{fig:RQ2-1h}
    \end{subfigure}
    \caption{Type I and Type II importance score obtained for families of code change properties over time, where all the values are normalised by total importance score of the given time.}
\label{fig:RQ2}
\end{figure*}

We calculated the Type I (mean decrease in accuracy) and the Type II (mean decrease in node impurity) importance score for each family of change code properties in case of both long and short period JIT models. The results shown in Figure \ref{fig:RQ2} are obtained for the testing period $j > 1$ using training data of period $i = j-1$ for short period JIT models and training data from periods $1$ to $i=j-1$ for long period models\footnote{We consider the index of period that are below the diagonal in Figure \ref{fig:RQ1-1}.}. To illustrate the importance score, we normalised them for each period; hence, the column-wise sum of all the values in Figure \ref{fig:RQ2} is 1. 

The overlaps of the graphs in Figure \ref{fig:RQ2} indicate that the rank of each family code property is changing by time. In some specific cases~(see Figure~\ref{fig:RQ1-1c} and \ref{fig:RQ1-1g}), there is almost no fluctuation in the importance ratio while the others are oscillating. Since there is no clear behaviour in the importance values, we can conclude that code change properties constantly evolve with the time. This finding is consistent with that of \citet{mcintosh2018} while they used Wald $\chi^2$ importance score for a nonlinear variant of multiple regression and we applied Type I and Type II importance score of Random Forest.

Here, unlike the original study, we can not assert that Size has the highest importance score across periods for both short- and long-period models. The importance rank of Size properties is changing especially in Figure~\ref{fig:RQ2-1d} and Figure~\ref{fig:RQ2-1h}. \\

These findings reveal that code change properties fluctuate across time horizons, suggesting that practitioners should retrain JIT models periodically.

\noindent\fcolorbox{black}{white}{%
    \minipage[t]{\dimexpr1\linewidth-2\fboxsep-2\fboxrule\relax}
        \textit{The importance score of the family of code change properties fluctuates as the project ages indicating the characteristics of fix-inducing changes have a propensity of varying from one period to another.}
    \endminipage}

\subsection{(RQ3) How can the performance of JIT models be improved while considering the chronology of the data?}

Two different strategies are hypothesised. First, we examine whether practitioners should utilise all the available data or only the recent data; second, we examine if weighted sampling (i.e. sampling more training instances from the recent data) can positively affect the performance of JIT models.

To test the first hypothesis, we carried out the Wilcoxon Signed Rank Test (WSRT), a non-parametric and distribution-free technique, to compare the predictive capabilities of two different models \cite{diebold1995}. In this test, the signs and the ranks of the predicted values are compared to identify whether the performance of two predictive models is different. 

\begin{table}[!ht]
\caption{Results of WSRT for six-month short- and long-period models, SPM and LPM, respectively}\label{tab:wsrt}
\centering
\begin{tabular}{c|ccc|ccc}
\toprule
 & SPM & LPM & $p$-value & SPM & LPM & $p$-value \\
 & \multicolumn{3}{c|}{AUC} & \multicolumn{3}{c}{BS} \\ \midrule
{\scshape JDT} & 0.73 & 0.75 & $0.002^{***}$ & 0.19 & 0.17 & $0.000^{***}$ \\ 
{\scshape Mozilla} & 0.80 & 0.82 & $0.003^{***}$ & 0.17 & 0.15 & $0.000^{***}$ \\ 
{\scshape Platform} & 0.77 & 0.78 & $0.004^{***}$ & 0.18 & 0.19 & $0.019^{***}$ \\ 
{\scshape Postgres} & 0.78 & 0.80 & $0.000^{***}$ & 0.17 & 0.19 & $0.000^{***}$ \\
\bottomrule
\end{tabular}
\end{table}

The null hypothesis states that there is no difference in the performance of JIT models using all the available data and the models using the recent data. The alternate hypothesis is the performance of JIT model using all data is better than that of using the recent data. We carried out one-tailed WSRT. 

From the results (refer Table \ref{tab:wsrt}), we can see that $p$-value is less than the significance level of $5\%$ ($\alpha= 0.05$), resulting in rejecting the null hypothesis. This indicates that using all the available data results in better performance of JIT models in all four projects. 


We tested the aforementioned hypotheses using the dataset from the original study. The results indicate that using recent dataset improves the performance of JIT models, which is contradictory to the results of our study.  
The results are publicly available at Github\footnote{https://github.com/HadiJahanshahi/JITChronology}. \\

Even though the performance of the model is significantly improved when we considered all the data, the impact of the size of training data also matters. As the size of the training data increases, so does the bias~\cite{Rahman:2013}, suggesting that the evolutionary properties (chronology) of code changes will be ignored if we only concentrate on the performance. Therefore, notwithstanding the improvement in the performance, we might be required to rely on the model trained on the recent dataset instead of using an all-inclusive model.

\noindent\fcolorbox{black}{white}{%
    \minipage[t]{\dimexpr1\linewidth-2\fboxsep-2\fboxrule\relax}
     \textit{The JIT models employing all the available data may outperform those using the recent data; therefore, the chronology of data does not affect the predictive power of JIT models. However, because of the evolution of code change properties (see RQ2), it is recommended to use the model obtained from more recent data.}
    \endminipage}


As the second hypothesis (and based on the conclusion of the original study), we examined whether sampling more data from the recent dataset can improve the predictive power of JIT models. Hence, as a threshold-independent approach, we randomly sampled $(1-\frac{i-k}{i})\times100\%$ of the changes for training period $i$, where, $k \in{\{1,\hdots,i\}}$. For instance, when the training period is $5$, we randomly sampled $(1-\frac{5-5}{5})\times100\% = 100\%$ of the changes from period $5$, $(1-\frac{5-4}{5})\times100\% = 80\%$ of the changes from period $4$, and so on. Later, we combined these changes and built a larger dataset that includes more data from the time period closer to the testing period. Using this approach, we determine whether the weighted sampling approach improves the accuracy of JIT prediction models. 

\begin{figure*}[!ht]
    \centering
    \begin{subfigure}[t]{0.4\textwidth}
        \centering
        \includegraphics[width=\textwidth]{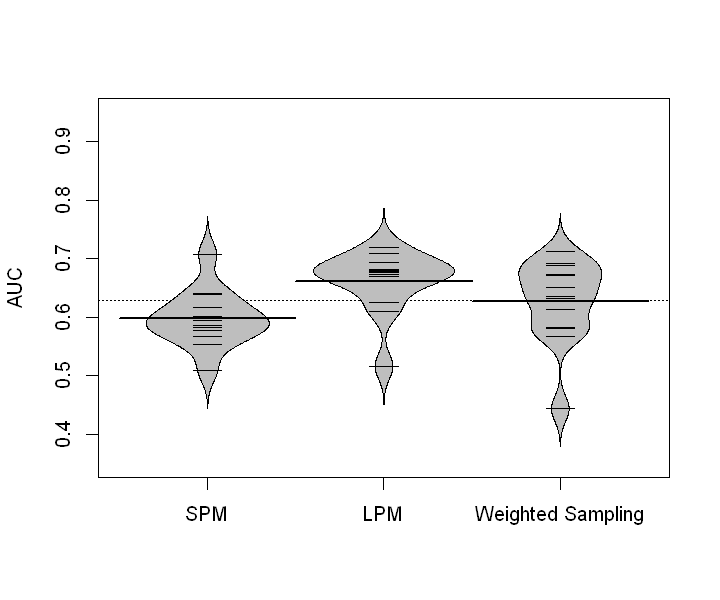}
        \vspace{-0.8cm}
        \captionsetup{justification=centering}
        \caption{AUC for ({\scshape JDT}) \\ $p$-value = 0.013}
        \label{fig:RQ4-1a}
    \end{subfigure}%
    ~
    \begin{subfigure}[t]{0.4\textwidth}
        \centering
        \includegraphics[width=\textwidth]{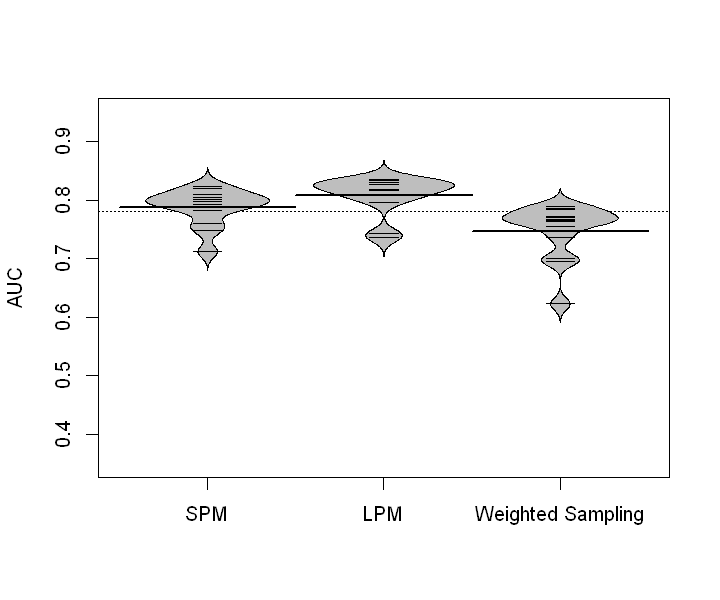}
        \vspace{-0.8cm}
        \captionsetup{justification=centering}
        \caption{AUC for ({\scshape Mozilla}) \\$p$-value = 0.000}
        \label{fig:RQ4-1b}
    \end{subfigure} 
    ~ 
    \begin{subfigure}[t]{0.4\textwidth}
        \centering
        \includegraphics[width=\textwidth]{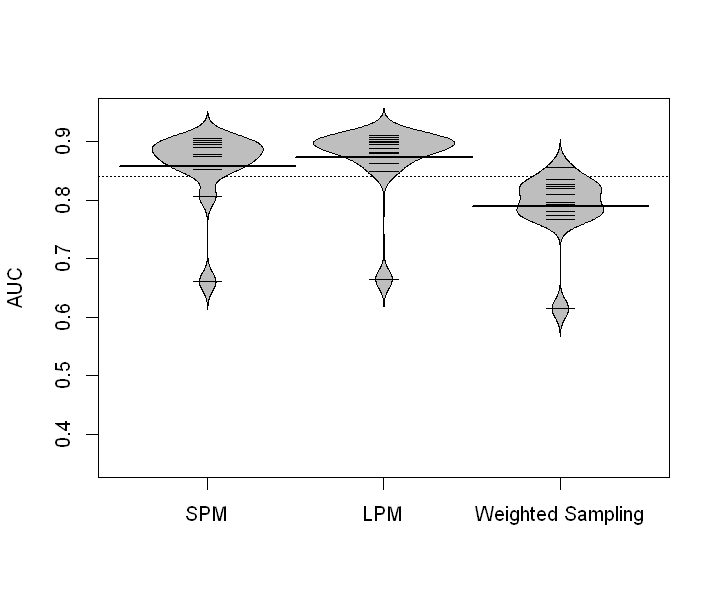}
        \vspace{-0.8cm}
        \captionsetup{justification=centering}
        \caption{AUC for ({\scshape Platform}) \\ $p$-value = 0.000}
        \label{fig:RQ4-1c}
    \end{subfigure}%
    ~ 
    \begin{subfigure}[t]{0.4\textwidth}
        \centering
        \includegraphics[width=\textwidth]{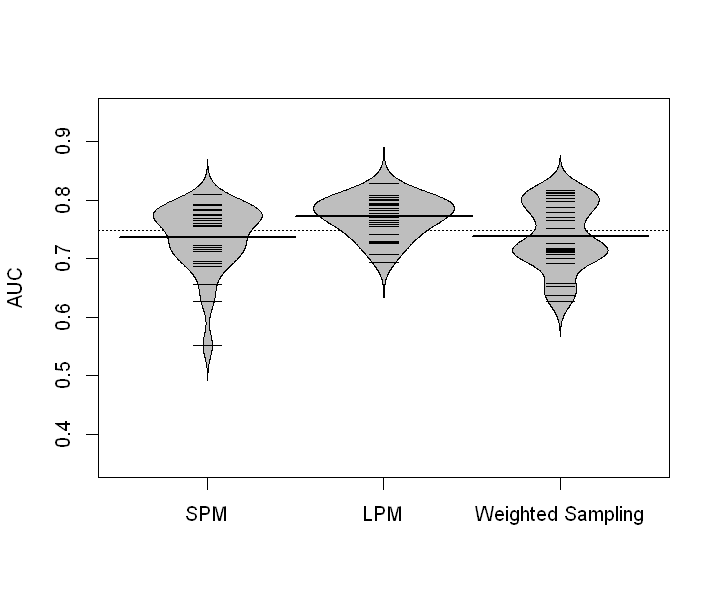}
        \vspace{-0.8cm}
        \captionsetup{justification=centering}
        \caption{AUC for ({\scshape Postgres}) \\ $p$-value = 0.002}
        \label{fig:RQ4-1d}
    \end{subfigure}
    
    \begin{subfigure}[t]{0.4\textwidth}
        \centering
        \includegraphics[width=\textwidth]{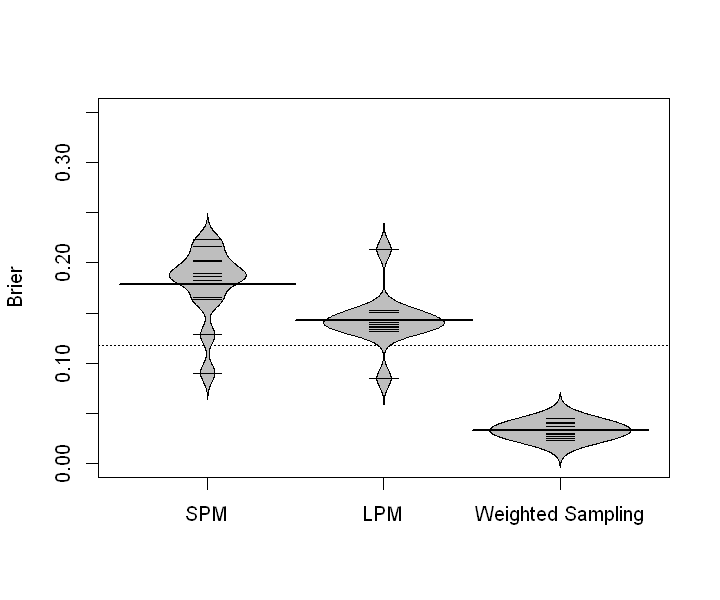}
        \vspace{-0.8cm}
        \captionsetup{justification=centering}
        \caption{Brier for ({\scshape JDT}) \\ $p$-value = 0.000}
        \label{fig:RQ4-1e}
    \end{subfigure}%
    ~
    \begin{subfigure}[t]{0.4\textwidth}
        \centering
        \includegraphics[width=\textwidth]{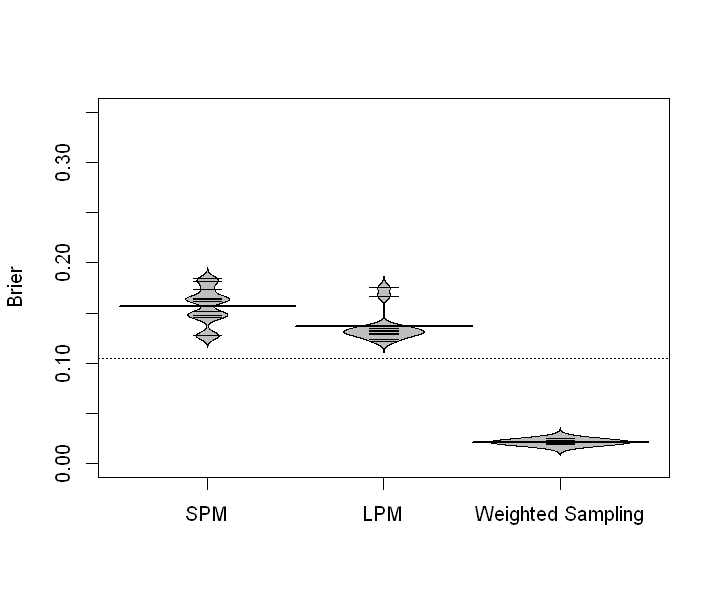}
        \vspace{-0.8cm}
        \captionsetup{justification=centering}
        \caption{Brier for ({\scshape Mozilla}) \\ $p$-value = 0.000}
        \label{fig:RQ4-1f}
    \end{subfigure}
    ~ 
    \begin{subfigure}[t]{0.4\textwidth}
        \centering
        \includegraphics[width=\textwidth]{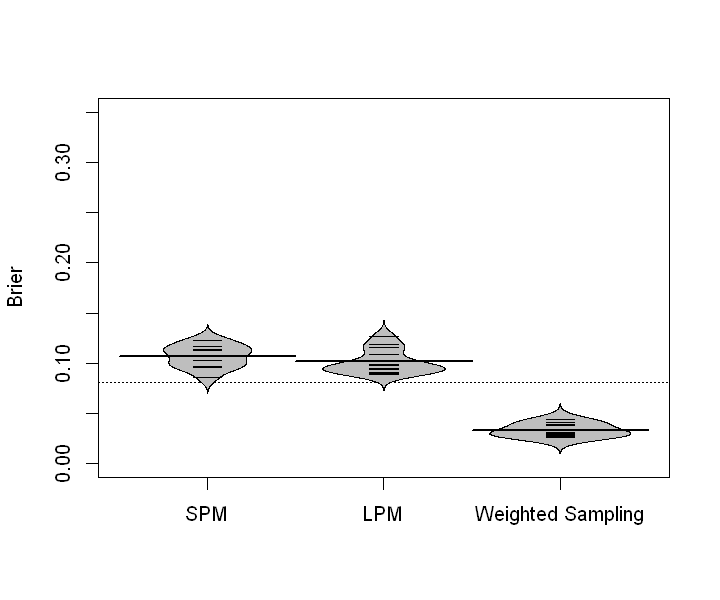}
        \vspace{-0.8cm}
        \captionsetup{justification=centering}
        \caption{Brier for ({\scshape Platform}) \\ $p$-value = 0.000}
        \label{fig:RQ4-1g}
    \end{subfigure}%
    ~ 
    \begin{subfigure}[t]{0.4\textwidth}
        \centering
        \includegraphics[width=\textwidth]{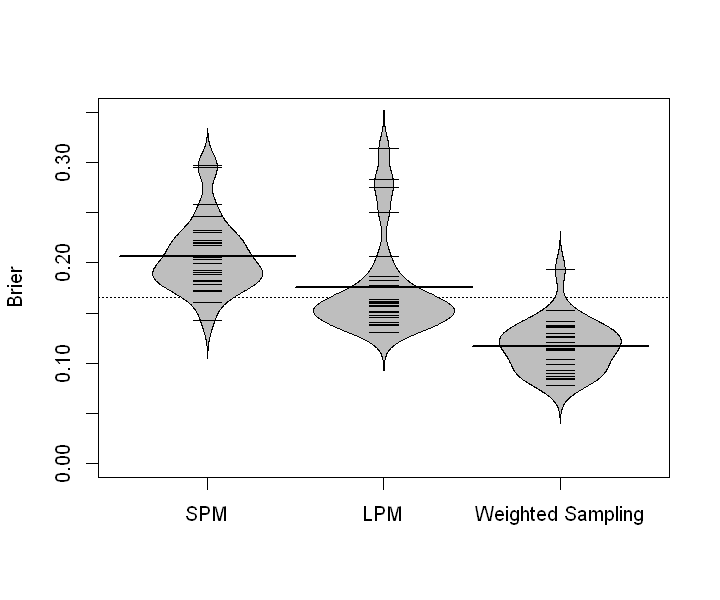}
        \vspace{-0.8cm}
        \captionsetup{justification=centering}
        \caption{Brier for ({\scshape Postgres}) \\ $p$-value = 0.000}
        \label{fig:RQ4-1h}
    \end{subfigure}
    
    \caption{AUC/Brier values for $j=n$ where $n$ is the number of period breakdowns}
\label{fig:RQ4}
\end{figure*}

Figure~\ref{fig:RQ4} demonstrates the AUC and BS values obtained for three different approaches, namely, $SPM$, $LPM$, and Weighted sampling, using beanplots. Similar to boxplots, a vertical curve in the beanplots denotes the distribution of the dataset and the long horizontal line specifies the median value. The plots are demonstrated for the testing dataset $j = n$ where $n$ is the total number of periods. For instance, the values of AUC for $SPM$ and $LPM$ in Figure~\ref{fig:RQ4-1a} are equivalent to the last column (Testing period $= 13$) of the heatmap in Figure~\ref{fig:RQ1-1a}. 

To check whether there is any significant difference in the median performance of three approaches, we used the Kruskal Wallis Test, a single-step multiple comparison procedure and a statistical test. The null hypothesis states that there is no difference in the median performances of JIT models using LPM, SPM and weighted sampling. The alternate hypothesis states that there is a statistically significant difference in the median performance of at least one of the JIT models. 

In all the cases, the p-value is less than the $\alpha = 0.05$, and we reject the null hypothesis. This indicates that there is a significant difference in the performance (AUC and BS) of JIT models using SPM, LPM, and weighted sampling.

On visual inspection, we can find that the performance (in terms of BS) of JIT models employing weighted sampling is better than LPM and SPM (refer Figure~\ref{fig:RQ4-1e} to Figure~\ref{fig:RQ4-1h}; however, considering AUC values for different strategies, we are unable to define the best one since the difference between AUC values is insignificant. \\


\noindent\fcolorbox{black}{white}{%
    \minipage[t]{\dimexpr1\linewidth-2\fboxsep-2\fboxrule\relax}
        \textit{There is a significant difference in the median performance among SPM-, LPM- and weighted sampling-based JIT models. In terms of Brier Score, the performance of the JIT models using weighted sampling is better than those using SPM and LPM. Hence, it is recommended that practitioners use the aforementioned weighted sampling approach to maintain the chronology of the data while achieving a better calibration score of the model.} 
        
    \endminipage}


\section{Threats to Validity}\label{sec:threatstovalidity}
Following are the threats to the validity of our study: 
\begin{enumerate}
    \item \textbf{Construct Validity:} The results obtained are dependent on the time window considered for the analysis. In this study, we considered a time period of six months consisting of sufficient volume of data to study the evolution of change code properties over time. 
    
    
    \item \textbf{Internal Validity:} 
    \begin{itemize}
        \item We used the dataset provided by \citet{kamei2013}. They have applied SZZ algorithm to identify the fix inducing changes. Despite its wide applicability, it suffers from several limitations such as incomplete mapping of the fixing commit and bug, inaccurate mapping as well as systematic bias caused by linking fix commit with no real bug report \cite{perez2018}. These limitations could affect the results of our study.
        
        \item The study considers only tangible factors related to the change code properties. Other intangible factors such as contributor turnover and changes to team culture are not considered. 
         
        \item The results of our study are valid only for classification models employing Random Forest and may not be applicable for other classification models such as Logistic Regression (LR). Nevertheless, the outcomes of some RQs using LR is accessible via GitHub page.
    \end{itemize}

    \item \textbf{External Validity:} We carried out the analysis on four open-source datasets with characteristics such as traceability and the presence of code review policy. Unlike the original study, the four datasets are not rapidly evolving. Hence, the results are not generalizable for all software systems. The study must be replicated on new datasets to study the fix-inducing code properties over time to generalise the results to the other JIT models.  
\end{enumerate}

\section{Conclusions across Studies}\label{sec:conclusion}
In this paper, we replicated \citet{mcintosh2018}'s work on understanding how the predictive power of JIT models changes over time. As an extension to their study, our work aims to study the effect of code change properties on JIT models over a time period and also, the effect of sampling approaches on the performance of JIT models. Based on the replication study and its extension, we can summarize our results as follows: 

\begin{enumerate}
    \item The predictive power of JIT models is not affected by time after being trained. Therefore, the predictive power of the model will not significantly change during the time horizon.
    \item The family importance scores of JIT models fluctuate as the system ages, indicating the change in the characteristic of datasets as time elapses.
    If we consider only the performance of the model, the chronology of the data does not play a significant role in model performance; however, the characteristic of the model which is trained on whole available data may not reflect the actual correlation between independent and dependent factors; thus, it is advised to build a model using the more recent data to both capture the evolution of the features and deliver an acceptable performance.
    \item The effect of fluctuations in the properties of fix-inducing changes with respect to the size of the data is considerable. Thus, the BS indicates that the suggested weighted sampling approach results in a better discriminatory power and calibration score; however, it is more complicated than other methods and needs constant training of the model. Thus, it is worthwhile to construct a model using the weighted sampling strategy to derive the benefits of both the satisfactory performance of training on all the accessible data and the chronological evolution of the data.
\end{enumerate}

Based on the results obtained and for better quality assurance plans, researchers and practitioners should take account of peaks and troughs in the importance of code change properties, thus training JIT models using a larger source of data, which contains the last six-month data.

Replication of the original study enabled us to confirm the impact of chronology on the importance of each family of code change properties. Furthermore, the outcome of RQ1 from our study was different than that of the original paper, indicating the need for further research to make practitioners confident about the generalizability of the issue. Finally, we proposed a new weighted sampling approach to complement the deduction made by the original study. The proposed approach is capable of augmenting the calibration ability of JIT models.

We used different methods and datasets to check the reliability and generalizability of the conclusion made by the original study. Hence, our conclusion may be restricted by the employed methodology or the quality of the datasets. For future research, this work can be replicated using systems that are developed in other contexts or the methods that are not covered by either of the works. To facilitate the replication of the paper, the codes that we used to analyze the RQs are available online\footnote{https://github.com/HadiJahanshahi/JITChronology}.


\bibliographystyle{ACM-Reference-Format}
\bibliography{ref}


\begin{thebibliography}{27}


\ifx \showCODEN    \undefined \def \showCODEN     #1{\unskip}     \fi
\ifx \showDOI      \undefined \def \showDOI       #1{#1}\fi
\ifx \showISBNx    \undefined \def \showISBNx     #1{\unskip}     \fi
\ifx \showISBNxiii \undefined \def \showISBNxiii  #1{\unskip}     \fi
\ifx \showISSN     \undefined \def \showISSN      #1{\unskip}     \fi
\ifx \showLCCN     \undefined \def \showLCCN      #1{\unskip}     \fi
\ifx \shownote     \undefined \def \shownote      #1{#1}          \fi
\ifx \showarticletitle \undefined \def \showarticletitle #1{#1}   \fi
\ifx \showURL      \undefined \def \showURL       {\relax}        \fi
\providecommand\bibfield[2]{#2}
\providecommand\bibinfo[2]{#2}
\providecommand\natexlab[1]{#1}
\providecommand\showeprint[2][]{arXiv:#2}

\bibitem[\protect\citeauthoryear{Breiman}{Breiman}{2001}]%
        {breiman2001}
\bibfield{author}{\bibinfo{person}{Leo Breiman}.}
  \bibinfo{year}{2001}\natexlab{}.
\newblock \showarticletitle{Random Forests}.
\newblock \bibinfo{journal}{\emph{Mach. Learn.}} \bibinfo{volume}{45},
  \bibinfo{number}{1} (\bibinfo{date}{Oct.} \bibinfo{year}{2001}),
  \bibinfo{pages}{5--32}.
\newblock
\showISSN{0885-6125}


\bibitem[\protect\citeauthoryear{{D'Ambros}, {Lanza}, and {Robbes}}{{D'Ambros}
  et~al\mbox{.}}{2010}]%
        {ambros2010}
\bibfield{author}{\bibinfo{person}{M. {D'Ambros}}, \bibinfo{person}{M.
  {Lanza}}, {and} \bibinfo{person}{R. {Robbes}}.}
  \bibinfo{year}{2010}\natexlab{}.
\newblock \showarticletitle{An extensive comparison of bug prediction
  approaches}. In \bibinfo{booktitle}{\emph{2010 7th IEEE Working Conference on
  Mining Software Repositories (MSR 2010)}}. \bibinfo{pages}{31--41}.
\newblock
\showISSN{2160-1852}
\urldef\tempurl%
\url{https://doi.org/10.1109/MSR.2010.5463279}
\showDOI{\tempurl}


\bibitem[\protect\citeauthoryear{Diebold and Mariano}{Diebold and
  Mariano}{1995}]%
        {diebold1995}
\bibfield{author}{\bibinfo{person}{Francis~X. Diebold} {and}
  \bibinfo{person}{Roberto~S. Mariano}.} \bibinfo{year}{1995}\natexlab{}.
\newblock \showarticletitle{Comparing Predictive Accuracy}.
\newblock \bibinfo{journal}{\emph{Journal of Business \& Economic Statistics}}
  \bibinfo{volume}{13}, \bibinfo{number}{3} (\bibinfo{year}{1995}),
  \bibinfo{pages}{253--263}.
\newblock


\bibitem[\protect\citeauthoryear{Giger, D'Ambros, Pinzger, and Gall}{Giger
  et~al\mbox{.}}{2012}]%
        {giger2012}
\bibfield{author}{\bibinfo{person}{Emanuel Giger}, \bibinfo{person}{Marco
  D'Ambros}, \bibinfo{person}{Martin Pinzger}, {and} \bibinfo{person}{Harald~C.
  Gall}.} \bibinfo{year}{2012}\natexlab{}.
\newblock \showarticletitle{Method-level Bug Prediction}. In
  \bibinfo{booktitle}{\emph{Proceedings of the ACM-IEEE International Symposium
  on Empirical Software Engineering and Measurement}}
  \emph{(\bibinfo{series}{ESEM '12})}. \bibinfo{publisher}{ACM},
  \bibinfo{address}{New York, NY, USA}, \bibinfo{pages}{171--180}.
\newblock
\showISBNx{978-1-4503-1056-7}
\urldef\tempurl%
\url{https://doi.org/10.1145/2372251.2372285}
\showDOI{\tempurl}


\bibitem[\protect\citeauthoryear{Gneiting and Raftery}{Gneiting and
  Raftery}{2007}]%
        {gneiting2007}
\bibfield{author}{\bibinfo{person}{Tilmann Gneiting} {and}
  \bibinfo{person}{Adrian~E Raftery}.} \bibinfo{year}{2007}\natexlab{}.
\newblock \showarticletitle{Strictly Proper Scoring Rules, Prediction, and
  Estimation}.
\newblock \bibinfo{journal}{\emph{J. Amer. Statist. Assoc.}}
  \bibinfo{volume}{102}, \bibinfo{number}{477} (\bibinfo{year}{2007}),
  \bibinfo{pages}{359--378}.
\newblock
\urldef\tempurl%
\url{https://doi.org/10.1198/016214506000001437}
\showDOI{\tempurl}
\showeprint{https://doi.org/10.1198/016214506000001437}


\bibitem[\protect\citeauthoryear{{Graves}, {Karr}, {Marron}, and
  {Siy}}{{Graves} et~al\mbox{.}}{2000}]%
        {graves2000}
\bibfield{author}{\bibinfo{person}{T.~L. {Graves}}, \bibinfo{person}{A.~F.
  {Karr}}, \bibinfo{person}{J.~S. {Marron}}, {and} \bibinfo{person}{H. {Siy}}.}
  \bibinfo{year}{2000}\natexlab{}.
\newblock \showarticletitle{Predicting fault incidence using software change
  history}.
\newblock \bibinfo{journal}{\emph{IEEE Transactions on Software Engineering}}
  \bibinfo{volume}{26}, \bibinfo{number}{7} (\bibinfo{date}{July}
  \bibinfo{year}{2000}), \bibinfo{pages}{653--661}.
\newblock
\showISSN{0098-5589}
\urldef\tempurl%
\url{https://doi.org/10.1109/32.859533}
\showDOI{\tempurl}


\bibitem[\protect\citeauthoryear{Guo, Zimmermann, Nagappan, and Murphy}{Guo
  et~al\mbox{.}}{2010}]%
        {guo2010}
\bibfield{author}{\bibinfo{person}{Philip~J. Guo}, \bibinfo{person}{Thomas
  Zimmermann}, \bibinfo{person}{Nachiappan Nagappan}, {and}
  \bibinfo{person}{Brendan Murphy}.} \bibinfo{year}{2010}\natexlab{}.
\newblock \showarticletitle{Characterizing and Predicting Which Bugs Get Fixed:
  An Empirical Study of Microsoft Windows}. In
  \bibinfo{booktitle}{\emph{Proceedings of the 32Nd ACM/IEEE International
  Conference on Software Engineering - Volume 1}} \emph{(\bibinfo{series}{ICSE
  '10})}. \bibinfo{publisher}{ACM}, \bibinfo{address}{New York, NY, USA},
  \bibinfo{pages}{495--504}.
\newblock
\showISBNx{978-1-60558-719-6}
\urldef\tempurl%
\url{https://doi.org/10.1145/1806799.1806871}
\showDOI{\tempurl}


\bibitem[\protect\citeauthoryear{Hand}{Hand}{2009}]%
        {hand2009}
\bibfield{author}{\bibinfo{person}{David~J. Hand}.}
  \bibinfo{year}{2009}\natexlab{}.
\newblock \showarticletitle{Measuring Classifier Performance: A Coherent
  Alternative to the Area Under the ROC Curve}.
\newblock \bibinfo{journal}{\emph{Mach. Learn.}} \bibinfo{volume}{77},
  \bibinfo{number}{1} (\bibinfo{date}{Oct.} \bibinfo{year}{2009}),
  \bibinfo{pages}{103--123}.
\newblock
\showISSN{0885-6125}
\urldef\tempurl%
\url{https://doi.org/10.1007/s10994-009-5119-5}
\showDOI{\tempurl}


\bibitem[\protect\citeauthoryear{{Hassan}}{{Hassan}}{2009}]%
        {hasan2009}
\bibfield{author}{\bibinfo{person}{A.~E. {Hassan}}.}
  \bibinfo{year}{2009}\natexlab{}.
\newblock \showarticletitle{Predicting faults using the complexity of code
  changes}. In \bibinfo{booktitle}{\emph{2009 IEEE 31st International
  Conference on Software Engineering}}. \bibinfo{pages}{78--88}.
\newblock
\showISSN{0270-5257}
\urldef\tempurl%
\url{https://doi.org/10.1109/ICSE.2009.5070510}
\showDOI{\tempurl}


\bibitem[\protect\citeauthoryear{Hata, Mizuno, and Kikuno}{Hata
  et~al\mbox{.}}{2012}]%
        {hata2012}
\bibfield{author}{\bibinfo{person}{Hideaki Hata}, \bibinfo{person}{Osamu
  Mizuno}, {and} \bibinfo{person}{Tohru Kikuno}.}
  \bibinfo{year}{2012}\natexlab{}.
\newblock \showarticletitle{Bug Prediction Based on Fine-grained Module
  Histories}. In \bibinfo{booktitle}{\emph{Proceedings of the 34th
  International Conference on Software Engineering}}
  \emph{(\bibinfo{series}{ICSE '12})}. \bibinfo{publisher}{IEEE Press},
  \bibinfo{address}{Piscataway, NJ, USA}, \bibinfo{pages}{200--210}.
\newblock
\showISBNx{978-1-4673-1067-3}


\bibitem[\protect\citeauthoryear{JD, J, A, KG, and A.}{JD
  et~al\mbox{.}}{2012}]%
        {malley2012}
\bibfield{author}{\bibinfo{person}{Malley JD}, \bibinfo{person}{Kruppa J},
  \bibinfo{person}{Dasgupta A}, \bibinfo{person}{Malley KG}, {and}
  \bibinfo{person}{Ziegler A.}} \bibinfo{year}{2012}\natexlab{}.
\newblock \showarticletitle{Probability machines: consistent probability
  estimation using nonparametric learning machines.}
\newblock \bibinfo{journal}{\emph{Methods of Information in Medicine}}
  \bibinfo{volume}{51}, \bibinfo{number}{1} (\bibinfo{year}{2012}),
  \bibinfo{pages}{74--81}.
\newblock


\bibitem[\protect\citeauthoryear{Kamei, Fukushima, Mcintosh, Yamashita,
  Ubayashi, and Hassan}{Kamei et~al\mbox{.}}{2016}]%
        {kamei2016}
\bibfield{author}{\bibinfo{person}{Yasutaka Kamei}, \bibinfo{person}{Takafumi
  Fukushima}, \bibinfo{person}{Shane Mcintosh}, \bibinfo{person}{Kazuhiro
  Yamashita}, \bibinfo{person}{Naoyasu Ubayashi}, {and}
  \bibinfo{person}{Ahmed~E. Hassan}.} \bibinfo{year}{2016}\natexlab{}.
\newblock \showarticletitle{Studying Just-in-time Defect Prediction Using
  Cross-project Models}.
\newblock \bibinfo{journal}{\emph{Empirical Softw. Engg.}}
  \bibinfo{volume}{21}, \bibinfo{number}{5} (\bibinfo{date}{Oct.}
  \bibinfo{year}{2016}), \bibinfo{pages}{2072--2106}.
\newblock
\showISSN{1382-3256}
\urldef\tempurl%
\url{https://doi.org/10.1007/s10664-015-9400-x}
\showDOI{\tempurl}


\bibitem[\protect\citeauthoryear{{Kamei}, {Matsumoto}, {Monden}, {Matsumoto},
  {Adams}, and {Hassan}}{{Kamei} et~al\mbox{.}}{2010}]%
        {kamei2010}
\bibfield{author}{\bibinfo{person}{Y. {Kamei}}, \bibinfo{person}{S.
  {Matsumoto}}, \bibinfo{person}{A. {Monden}}, \bibinfo{person}{K.
  {Matsumoto}}, \bibinfo{person}{B. {Adams}}, {and} \bibinfo{person}{A.~E.
  {Hassan}}.} \bibinfo{year}{2010}\natexlab{}.
\newblock \showarticletitle{Revisiting common bug prediction findings using
  effort-aware models}. In \bibinfo{booktitle}{\emph{2010 IEEE International
  Conference on Software Maintenance}}. \bibinfo{pages}{1--10}.
\newblock
\showISSN{1063-6773}
\urldef\tempurl%
\url{https://doi.org/10.1109/ICSM.2010.5609530}
\showDOI{\tempurl}


\bibitem[\protect\citeauthoryear{{Kamei}, {Shihab}, {Adams}, {Hassan},
  {Mockus}, {Sinha}, and {Ubayashi}}{{Kamei} et~al\mbox{.}}{2013}]%
        {kamei2013}
\bibfield{author}{\bibinfo{person}{Y. {Kamei}}, \bibinfo{person}{E. {Shihab}},
  \bibinfo{person}{B. {Adams}}, \bibinfo{person}{A.~E. {Hassan}},
  \bibinfo{person}{A. {Mockus}}, \bibinfo{person}{A. {Sinha}}, {and}
  \bibinfo{person}{N. {Ubayashi}}.} \bibinfo{year}{2013}\natexlab{}.
\newblock \showarticletitle{A large-scale empirical study of just-in-time
  quality assurance}.
\newblock \bibinfo{journal}{\emph{IEEE Transactions on Software Engineering}}
  \bibinfo{volume}{39}, \bibinfo{number}{6} (\bibinfo{date}{June}
  \bibinfo{year}{2013}), \bibinfo{pages}{757--773}.
\newblock
\showISSN{0098-5589}
\urldef\tempurl%
\url{https://doi.org/10.1109/TSE.2012.70}
\showDOI{\tempurl}


\bibitem[\protect\citeauthoryear{{Kim}, {Whitehead, Jr.}, and {Zhang}}{{Kim}
  et~al\mbox{.}}{2008}]%
        {kim2008}
\bibfield{author}{\bibinfo{person}{S. {Kim}}, \bibinfo{person}{E.~J.
  {Whitehead, Jr.}}, {and} \bibinfo{person}{Y. {Zhang}}.}
  \bibinfo{year}{2008}\natexlab{}.
\newblock \showarticletitle{Classifying Software Changes: Clean or Buggy?}
\newblock \bibinfo{journal}{\emph{IEEE Transactions on Software Engineering}}
  \bibinfo{volume}{34}, \bibinfo{number}{2} (\bibinfo{date}{March}
  \bibinfo{year}{2008}), \bibinfo{pages}{181--196}.
\newblock
\showISSN{0098-5589}
\urldef\tempurl%
\url{https://doi.org/10.1109/TSE.2007.70773}
\showDOI{\tempurl}


\bibitem[\protect\citeauthoryear{{Koru}, {Zhang}, {El Emam}, and {Liu}}{{Koru}
  et~al\mbox{.}}{2009}]%
        {koru2009}
\bibfield{author}{\bibinfo{person}{A.~G. {Koru}}, \bibinfo{person}{D. {Zhang}},
  \bibinfo{person}{K. {El Emam}}, {and} \bibinfo{person}{H. {Liu}}.}
  \bibinfo{year}{2009}\natexlab{}.
\newblock \showarticletitle{An Investigation into the Functional Form of the
  Size-Defect Relationship for Software Modules}.
\newblock \bibinfo{journal}{\emph{IEEE Transactions on Software Engineering}}
  \bibinfo{volume}{35}, \bibinfo{number}{2} (\bibinfo{date}{March}
  \bibinfo{year}{2009}), \bibinfo{pages}{293--304}.
\newblock
\showISSN{0098-5589}
\urldef\tempurl%
\url{https://doi.org/10.1109/TSE.2008.90}
\showDOI{\tempurl}


\bibitem[\protect\citeauthoryear{Li, Herbsleb, Shaw, and Robinson}{Li
  et~al\mbox{.}}{2006}]%
        {li2006}
\bibfield{author}{\bibinfo{person}{Paul~Luo Li}, \bibinfo{person}{James
  Herbsleb}, \bibinfo{person}{Mary Shaw}, {and} \bibinfo{person}{Brian
  Robinson}.} \bibinfo{year}{2006}\natexlab{}.
\newblock \showarticletitle{Experiences and Results from Initiating Field
  Defect Prediction and Product Test Prioritization Efforts at ABB Inc.}. In
  \bibinfo{booktitle}{\emph{Proceedings of the 28th International Conference on
  Software Engineering}} \emph{(\bibinfo{series}{ICSE '06})}.
  \bibinfo{publisher}{ACM}, \bibinfo{address}{New York, NY, USA},
  \bibinfo{pages}{413--422}.
\newblock
\showISBNx{1-59593-375-1}
\urldef\tempurl%
\url{https://doi.org/10.1145/1134285.1134343}
\showDOI{\tempurl}


\bibitem[\protect\citeauthoryear{Matsumoto, Kamei, Monden, Matsumoto, and
  Nakamura}{Matsumoto et~al\mbox{.}}{2010}]%
        {matsumoto2010}
\bibfield{author}{\bibinfo{person}{Shinsuke Matsumoto},
  \bibinfo{person}{Yasutaka Kamei}, \bibinfo{person}{Akito Monden},
  \bibinfo{person}{Ken-ichi Matsumoto}, {and} \bibinfo{person}{Masahide
  Nakamura}.} \bibinfo{year}{2010}\natexlab{}.
\newblock \showarticletitle{An Analysis of Developer Metrics for Fault
  Prediction}. In \bibinfo{booktitle}{\emph{Proceedings of the 6th
  International Conference on Predictive Models in Software Engineering}}
  \emph{(\bibinfo{series}{PROMISE '10})}. \bibinfo{publisher}{ACM},
  \bibinfo{address}{New York, NY, USA}, Article \bibinfo{articleno}{18},
  \bibinfo{numpages}{9}~pages.
\newblock
\showISBNx{978-1-4503-0404-7}
\urldef\tempurl%
\url{https://doi.org/10.1145/1868328.1868356}
\showDOI{\tempurl}


\bibitem[\protect\citeauthoryear{{McIntosh} and {Kamei}}{{McIntosh} and
  {Kamei}}{2018}]%
        {mcintosh2018}
\bibfield{author}{\bibinfo{person}{S. {McIntosh}} {and} \bibinfo{person}{Y.
  {Kamei}}.} \bibinfo{year}{2018}\natexlab{}.
\newblock \showarticletitle{Are Fix-Inducing Changes a Moving Target? A
  Longitudinal Case Study of Just-In-Time Defect Prediction}.
\newblock \bibinfo{journal}{\emph{IEEE Transactions on Software Engineering}}
  \bibinfo{volume}{44}, \bibinfo{number}{5} (\bibinfo{date}{May}
  \bibinfo{year}{2018}), \bibinfo{pages}{412--428}.
\newblock
\showISSN{0098-5589}
\urldef\tempurl%
\url{https://doi.org/10.1109/TSE.2017.2693980}
\showDOI{\tempurl}


\bibitem[\protect\citeauthoryear{{Mockus} and {Weiss}}{{Mockus} and
  {Weiss}}{2000}]%
        {mockus2000}
\bibfield{author}{\bibinfo{person}{A. {Mockus}} {and} \bibinfo{person}{D.~M.
  {Weiss}}.} \bibinfo{year}{2000}\natexlab{}.
\newblock \showarticletitle{Predicting risk of software changes}.
\newblock \bibinfo{journal}{\emph{Bell Labs Technical Journal}}
  \bibinfo{volume}{5}, \bibinfo{number}{2} (\bibinfo{date}{April}
  \bibinfo{year}{2000}), \bibinfo{pages}{169--180}.
\newblock
\showISSN{1538-7305}
\urldef\tempurl%
\url{https://doi.org/10.1002/bltj.2229}
\showDOI{\tempurl}


\bibitem[\protect\citeauthoryear{Nagappan and Ball}{Nagappan and Ball}{2005}]%
        {nagappan2005}
\bibfield{author}{\bibinfo{person}{Nachiappan Nagappan} {and}
  \bibinfo{person}{Thomas Ball}.} \bibinfo{year}{2005}\natexlab{}.
\newblock \showarticletitle{Use of Relative Code Churn Measures to Predict
  System Defect Density}. In \bibinfo{booktitle}{\emph{Proceedings of the 27th
  International Conference on Software Engineering}}
  \emph{(\bibinfo{series}{ICSE '05})}. \bibinfo{publisher}{ACM},
  \bibinfo{address}{New York, NY, USA}, \bibinfo{pages}{284--292}.
\newblock
\showISBNx{1-58113-963-2}
\urldef\tempurl%
\url{https://doi.org/10.1145/1062455.1062514}
\showDOI{\tempurl}


\bibitem[\protect\citeauthoryear{Nagappan, Ball, and Zeller}{Nagappan
  et~al\mbox{.}}{2006}]%
        {nagappan2006}
\bibfield{author}{\bibinfo{person}{Nachiappan Nagappan},
  \bibinfo{person}{Thomas Ball}, {and} \bibinfo{person}{Andreas Zeller}.}
  \bibinfo{year}{2006}\natexlab{}.
\newblock \showarticletitle{Mining Metrics to Predict Component Failures}. In
  \bibinfo{booktitle}{\emph{Proceedings of the 28th International Conference on
  Software Engineering}} \emph{(\bibinfo{series}{ICSE '06})}.
  \bibinfo{publisher}{ACM}, \bibinfo{address}{New York, NY, USA},
  \bibinfo{pages}{452--461}.
\newblock
\showISBNx{1-59593-375-1}
\urldef\tempurl%
\url{https://doi.org/10.1145/1134285.1134349}
\showDOI{\tempurl}


\bibitem[\protect\citeauthoryear{Rahman, Posnett, Herraiz, and Devanbu}{Rahman
  et~al\mbox{.}}{2013}]%
        {Rahman:2013}
\bibfield{author}{\bibinfo{person}{Foyzur Rahman}, \bibinfo{person}{Daryl
  Posnett}, \bibinfo{person}{Israel Herraiz}, {and} \bibinfo{person}{Premkumar
  Devanbu}.} \bibinfo{year}{2013}\natexlab{}.
\newblock \showarticletitle{Sample Size vs. Bias in Defect Prediction}. In
  \bibinfo{booktitle}{\emph{Proceedings of the 2013 9th Joint Meeting on
  Foundations of Software Engineering}} \emph{(\bibinfo{series}{ESEC/FSE
  2013})}. \bibinfo{publisher}{ACM}, \bibinfo{address}{New York, NY, USA},
  \bibinfo{pages}{147--157}.
\newblock
\showISBNx{978-1-4503-2237-9}
\urldef\tempurl%
\url{https://doi.org/10.1145/2491411.2491418}
\showDOI{\tempurl}


\bibitem[\protect\citeauthoryear{Rodr{\'i}guez-P{\'e}rez, Robles, and
  Gonz{\'a}lez-Barahona}{Rodr{\'i}guez-P{\'e}rez et~al\mbox{.}}{2018}]%
        {perez2018}
\bibfield{author}{\bibinfo{person}{Gema Rodr{\'i}guez-P{\'e}rez},
  \bibinfo{person}{Gregorio Robles}, {and} \bibinfo{person}{Jes{\'u}s~M.
  Gonz{\'a}lez-Barahona}.} \bibinfo{year}{2018}\natexlab{}.
\newblock \showarticletitle{Reproducibility and credibility in empirical
  software engineering: A case study based on a systematic literature review of
  the use of the SZZ algorithm}.
\newblock \bibinfo{journal}{\emph{Information and Software Technology}}
  \bibinfo{volume}{99} (\bibinfo{year}{2018}), \bibinfo{pages}{164 -- 176}.
\newblock
\showISSN{0950-5849}
\urldef\tempurl%
\url{https://doi.org/10.1016/j.infsof.2018.03.009}
\showDOI{\tempurl}


\bibitem[\protect\citeauthoryear{{Shaoqing Ren}, {Cao}, {Yichen Wei}, and
  {Sun}}{{Shaoqing Ren} et~al\mbox{.}}{2015}]%
        {ren2015}
\bibfield{author}{\bibinfo{person}{{Shaoqing Ren}}, \bibinfo{person}{X. {Cao}},
  \bibinfo{person}{{Yichen Wei}}, {and} \bibinfo{person}{J. {Sun}}.}
  \bibinfo{year}{2015}\natexlab{}.
\newblock \showarticletitle{Global refinement of random forest}. In
  \bibinfo{booktitle}{\emph{2015 IEEE Conference on Computer Vision and Pattern
  Recognition (CVPR)}}. \bibinfo{pages}{723--730}.
\newblock
\showISSN{1063-6919}
\urldef\tempurl%
\url{https://doi.org/10.1109/CVPR.2015.7298672}
\showDOI{\tempurl}


\bibitem[\protect\citeauthoryear{Strobl, Boulesteix, Zeileis, and
  Hothorn}{Strobl et~al\mbox{.}}{2007}]%
        {strobl2007}
\bibfield{author}{\bibinfo{person}{Carolin Strobl}, \bibinfo{person}{Anne-Laure
  Boulesteix}, \bibinfo{person}{Achim Zeileis}, {and} \bibinfo{person}{Torsten
  Hothorn}.} \bibinfo{year}{2007}\natexlab{}.
\newblock \showarticletitle{Bias in random forest variable importance measures:
  illustrations, sources and a solution.}
\newblock \bibinfo{journal}{\emph{BMC Bioinformatics}} \bibinfo{volume}{8},
  \bibinfo{number}{25} (\bibinfo{date}{Oct.} \bibinfo{year}{2007}),
  \bibinfo{pages}{1--25}.
\newblock


\bibitem[\protect\citeauthoryear{Zimmermann, Premraj, and Zeller}{Zimmermann
  et~al\mbox{.}}{2007}]%
        {zimmermann2007}
\bibfield{author}{\bibinfo{person}{Thomas Zimmermann}, \bibinfo{person}{Rahul
  Premraj}, {and} \bibinfo{person}{Andreas Zeller}.}
  \bibinfo{year}{2007}\natexlab{}.
\newblock \showarticletitle{Predicting Defects for Eclipse}. In
  \bibinfo{booktitle}{\emph{Proceedings of the Third International Workshop on
  Predictor Models in Software Engineering}} \emph{(\bibinfo{series}{PROMISE
  '07})}. \bibinfo{publisher}{IEEE Computer Society},
  \bibinfo{address}{Washington, DC, USA}, \bibinfo{pages}{9--}.
\newblock
\showISBNx{0-7695-2954-2}
\urldef\tempurl%
\url{https://doi.org/10.1109/PROMISE.2007.10}
\showDOI{\tempurl}


\end{thebibliography}


\end{document}